\documentclass[showpacs,showkeys,amsmath,amssymb,superscriptaddress,reprint, preprintnumbers,nofootinbib,notitlepage, groupdaddress,showabstract,aps,10pt,prd,notitlepage,showpacs,nofootinbib,superscriptaddress,compressed]{revtex4-1}

\pdfoutput=1 % if your are submitting a pdflatex (i.e. if you have
\usepackage[font=small]{caption}
\usepackage[utf8]{inputenc}
\usepackage[english]{babel}
\usepackage{bm}
\usepackage{graphicx}
\usepackage{dcolumn}
\usepackage{pbox}
\usepackage{epsfig}
\usepackage[dvipsnames]{xcolor}
\usepackage{slashed}
\usepackage{amssymb}
\usepackage{mathrsfs}
\usepackage{color}
\usepackage[font=small]{caption}
\usepackage[font=small]{subcaption}
\usepackage{url}
\usepackage{url}
\usepackage{braket}
\usepackage{hyperref}
\usepackage{fontawesome5}
\definecolor{DarkBlue}{rgb}{0.7, 0.4, 1}
\definecolor{Blue}{rgb}{0, 0.8, 0}
\definecolor{MyLightBlue}{rgb}{0.5,0.7,1.9}
\definecolor{MyGreen}{rgb}{0.0,0.2, 0.0}
\definecolor{MyBrickRed}{rgb}{0, 0.5, 0.2}
\RequirePackage{hyperref}
\hypersetup{colorlinks, citecolor=Blue,linkcolor=DarkBlue, urlcolor=Green}
\usepackage[normalem]{ulem}

\newcommand{\bea}{\begin{eqnarray}}
\newcommand{\eea}{\end{eqnarray}}

\makeatletter

\renewcommand\@makecaption[2]{%
  \par
  \vskip\abovecaptionskip
  \begingroup

   \small\rmfamily
    \begingroup
     \samepage
     \flushing
     \let\footnote\@footnotemark@gobble
     \@make@capt@title{#1}{#2}\par
    \endgroup
  \endgroup
  \vskip\belowcaptionskip
}
\makeatother

\DeclareUnicodeCharacter{2212}{-}
\newcommand{\SM}{\text{SM}}

\begin{document}
%%%%%%%%%%%%%%%%%%%%%%%%%%%%%%
\title{Heavy neutral leptons from light scalar in fixed target and forward search experiments}
%%%%%%%%%%%%%%%%%%%%%
\author{ShivaSankar K.A.}\email{a-shiva@particle.sci.hokudai.ac.jp}
\affiliation{Department of Physics, Hokkaido University, Sapporo 060-0810, Japan}
\author{Souvik Das}\email{souvik\_das@tamu.edu}
\affiliation{Department of Physics and Astronomy, Mitchell Institute for Fundamental Physics and Astronomy, Texas A\&M University, College Station, Texas 77843, USA}
\author{Arindam Das}
\email{adas@particle.sci.hokudai.ac.jp}
\affiliation{Institute for the Advancement of Higher Education, Hokkaido University, Sapporo 060-0817, Japan}
\affiliation{Department of Physics, Hokkaido University, Sapporo 060-0810, Japan}
\author{Sanjoy Mandal}
\email{smandal@kias.re.kr}
\affiliation{Korea Institute for Advanced Study, Seoul 02455, Korea}
%%%%%%%%%%%%%%%%%%%%%%%%%%%
\begin{abstract}
%%%%%%%%%%%%%%%%%%%%%%%%%%%%%%%%%%%%%
The observation of neutrino masses strongly motivates $U(1)_{B-L}$ extensions of the Standard Model, in which heavy neutral leptons acquire Majorana masses through spontaneous $U(1)_{B-L}$ symmetry breaking and generate light neutrino masses via the seesaw mechanism. In this framework, the singlet scalar responsible for symmetry breaking mixes with the SM Higgs boson, allowing it to be produced in rare meson decays. We investigate a scenario in which this light scalar promptly decays into a pair of long-lived heavy neutrinos that subsequently decay into visible charged leptons and hadrons through light-heavy neutrino mixing inside the proposed Forward Physics Facility (FPF) at the FCC-hh and the SHiP beam-dump experiment. Taking into account realistic detector geometries, decay probabilities, and visible branching fractions, we estimate the projected sensitivities to the scalar-Higgs mixing angle as a function of the scalar mass and to the light-heavy neutrino mixing as a function of the heavy neutrino mass. We find that FPF and SHiP can significantly extend the discovery reach for both light scalars and long-lived heavy neutrinos beyond existing experimental limits, providing powerful and complementary probes of neutrino-mass generation and hidden-sector physics.\href{https://github.com/SouvikPhD/RHN-Detection-with-FASER-2-}{\faGithub}
%%%%%%%%%%%%%%%%%%%%%%%%%%%%%%%%%%%%%%
\end{abstract}
%%%%%%%%%%%%%%%%%%%%%%%%%%%
\maketitle
%%%%%%%%%%%%%%%%%%%%%%%%%%%%%%%%%%%%%%%%%%%%%%%%%%
\noindent

%%%%%%%%main text%%%%%%%%%%%%%%%%%%%%%%%%%%%%%%%%%
\section{Introduction}
%%%%%%%%%%%%%%%%%%%%%%%%%%%%%%%%%%%%%%%%%%%%%%%%
The discovery of neutrino masses and flavor mixing~\cite{ParticleDataGroup:2020ssz}, phenomena that cannot be explained within the Standard Model (SM), provides compelling evidence for physics beyond the SM~(BSM). One of the most elegant and minimal frameworks addressing this puzzle is the seesaw mechanism~\cite{Minkowski:1977sc,Yanagida:1979as,Gell-Mann:1979vob,Mohapatra:1979ia,Schechter:1980gr}, which extends the SM by introducing heavy gauge-singlet neutral leptons and naturally explains the observed smallness of active neutrino masses.

Motivated by this framework, we consider an anomaly-free $U(1)_{\rm B-L}$ gauge extension of the SM~\cite{Davidson:1978pm,Davidson:1979wr,Marshak:1979fm,Mohapatra:1980qe}. The model contains three generations of right-handed neutrinos (RHNs) together with an SM-singlet scalar responsible for the spontaneous breaking of the $U(1)_{\rm B-L}$ symmetry. Once the singlet scalar acquires a non-zero vacuum expectation value (VEV), it generates Majorana masses for the RHNs. Following electroweak symmetry breaking, the seesaw mechanism gives rise to the observed light neutrino masses and flavor mixing. Consequently, the heavy neutrinos interact with the SM particles through their suppressed mixing with the light neutrinos, while coupling directly to the new $Z^\prime$ gauge boson owing to their non-vanishing B$-$L charge.

The scalar sector contains an additional SM-singlet scalar that mixes with the SM Higgs boson through the renormalizable scalar potential. This mixing induces its interactions with SM particles, while its Yukawa couplings allow it to decay into a pair of heavy neutrinos whenever kinematically accessible. In this work, we investigate the scenario in which a light, short-lived singlet scalar is produced via rare meson decays and subsequently decays predominantly into a pair of long-lived heavy neutrinos. These heavy neutrinos travel macroscopic distances before decaying into visible final states inside the detector, leading to distinctive LLP signatures. This is complementary to the conventional LLP scenario, where the scalar itself is long-lived and decays into SM particles. To the best of our knowledge, this is the first dedicated study of LLP signatures arising from promptly decaying singlet scalars that produce long-lived heavy neutrinos.

To explore the discovery potential of this scenario, we consider two complementary future intensity-frontier facilities. The first is the proposed Search for Hidden Particles (SHiP) experiment~\cite{Alekhin:2015byh, SHiP:2020vbd,SHiP:2021nfo,Experiments:2845542}, a proton beam-dump facility designed to search for feebly interacting particles. SHiP is planned to employ a $50~\mathrm{m}$ long trapezoidal decay volume with transverse dimensions of $5~\mathrm{m}\times10~\mathrm{m}$ at the upstream end and $6~\mathrm{m}\times12~\mathrm{m}$ at the downstream end, located approximately $70~\mathrm{m}$ downstream of a titanium-zirconium-molybdenum (TZM) target. Using a $400~\mathrm{GeV}$ proton beam from the CERN SPS, SHiP aims to accumulate $4\times10^{19}$ protons on target (PoT) per year~\cite{Bartosik:2650722}, corresponding to a projected total of approximately $6\times10^{20}$ PoT over 15 years.

We also consider the proposed Forward Physics Facility (FPF) at the Future Circular Collider operating in the hadron-hadron mode (FCC-hh)~\cite{FCC:2018byv,FCC:2018vvp,Aleksa:2019pvl,MammenAbraham:2024gun}. With a center-of-mass energy of $\sqrt{s}=100$~TeV and an integrated luminosity of up to $30~\mathrm{ab}^{-1}$, the FCC-hh will produce an enormous flux of particles in the far-forward region, making it an ideal environment for LLP searches. Located approximately $1.5~\mathrm{km}$ downstream of the interaction point, the FPF is designed to detect visible decays of LLPs while employing a dedicated sweeper magnet to suppress charged-particle backgrounds. Following Ref.~\cite{MammenAbraham:2024gun}, we consider two detector configurations. The first, FPF1, consists of a $5\,\mathrm{m}\times5\,\mathrm{m}\times50\,\mathrm{m}$ decay volume, whereas the second, FPF2, employs a significantly larger $20\,\mathrm{m}\times20\,\mathrm{m}\times400\,\mathrm{m}$ decay volume, assuming a visible-energy threshold of $E_{\rm vis}\ge100$~GeV. Both detector concepts feature magnetic spectrometers for precise charged-particle reconstruction, veto systems to reject incoming charged particles, a vacuum decay volume to suppress neutrino-induced backgrounds, and a downstream electromagnetic calorimeter for efficient reconstruction of electrons and positrons from LLP decays.

In this paper, we investigate the production of a light, short-lived singlet scalar through rare meson decays within the $U(1)_{\rm B-L}$ framework at the SHiP and FPF experiments. The scalar subsequently decays into a pair of long-lived heavy neutrinos, which propagate to the detector and decay into visible final states. By analyzing these signatures, we derive the projected sensitivities to the scalar-Higgs mixing angle as a function of the singlet scalar mass and to the light-heavy neutrino mixing as a function of the heavy neutrino mass. Our study demonstrates that future beam-dump and far-forward experiments provide powerful and complementary probes of the scalar and neutrino sectors of the minimal $U(1)_{\rm B-L}$ model.

The remainder of this paper is organized as follows. In Sec.~\ref{secII}, we briefly review the $U(1)_{\rm B-L}$ model and its relevant interactions. The phenomenological analysis and projected sensitivities are presented in Sec.~\ref{secIII}. Finally, we summarize our conclusions in Sec.~\ref{secIV}.
%%%%%%%%%%%%%%%%%%%
\section{The framework}
\label{secII}
%%%%%%%%%%%%%%%%%%%%%%%%%%%%%%%%%%%%%%%%%%%%%%%%%%%%%%%
Under the $\mathrm{SM}\otimes U(1)_{\rm B-L}$ gauge symmetry, the SM quark fields transform as $q_L^i=(3,2,\frac{1}{6},\frac{1}{3})$, $u_R^i=(3,1,\frac{2}{3},\frac{1}{3})$, and $d_R^i=(3,1,-\frac{1}{3},\frac{1}{3})$, where $i=1,2,3$ denotes the generation index. The SM lepton fields transform as $\ell_L^i=(1,2,-\frac{1}{2},-1)$ and $e_R^i=(1,1,-1,-1)$, while the SM Higgs doublet is assigned the quantum numbers $H=(1,2,\frac{1}{2},0)$.

To render the theory free from gauge and mixed gauge-gravitational anomalies, we need to introduce three generations of SM-singlet RHNs, transforming as $N_R^i=(1,1,0,-1)$ with $i=1,2,3$. The model also contains an SM-singlet scalar field, $\Phi=(1,1,0,2)$, whose non-zero VEV spontaneously breaks the $U(1)_{\rm B-L}$ gauge symmetry. This symmetry breaking generates Majorana masses for the RHNs through Yukawa interactions. Following electroweak symmetry breaking, the neutrino Dirac mass terms are generated, and the interplay between the Dirac and Majorana mass matrices realizes the type-I seesaw mechanism, naturally explaining the observed smallness of light neutrino masses and their flavor mixing.

The relevant Yukawa interactions for the B$-$L scenario can be written as,
\bea
{\cal L} &\supset& - Y_{\nu_{\alpha \beta}} \overline{\ell_L^\alpha} \tilde{H}\,N_R^\beta- \frac{1}{2}Y_{N_\alpha} \Phi \overline{(N_R^\alpha)^c} N_R^\alpha + {\rm H.c.,}
\label{eq:LYk}
\eea
where we start-off in a basis where the $Y_{N_{\alpha}}$ matrix is diagonal, and $\tilde{H} = i \tau^2 H^*$ with $\tau^2$ being the second Pauli matrix. The scalar potential involving two scalar fields is given by
\bea
  V \ &=& -\ m_H^2(H^\dag H) + \lambda_H^{} (H^\dag H)^2 - m_\Phi^2 (\Phi^\dag \Phi) + \lambda_\Phi^{} (\Phi^\dag \Phi)^2 \nonumber \\
  &&+\lambda_{\rm mix} (H^\dag H)(\Phi^\dag \Phi)~.
\eea
After the B$-$L and electroweak gauge symmetries are broken, the VEVs of $H$ and $\Phi$ are developed following
\begin{align}
%\bea
\label{eq:VEV}
  \langle H\rangle \ = \ \frac{1}{\sqrt{2}}\begin{pmatrix} v+h\\0
  \end{pmatrix}~, \quad {\rm and}\quad
  \langle\Phi\rangle \ =\  \frac{v_{\rm B-L}^{}+\phi}{\sqrt{2}}~,
\end{align}
%\eea
where electroweak scale is set at $v=246$ GeV. In our analysis we consider $v_{\rm B-L}\gg v$ following LEP bound as $v_{\rm B-L} > 3.5$ TeV from \cite{KA:2023dyz}. After B$-$L symmetry breaking, the mass of neutral, BSM gauge boson $Z^\prime$ is generated as
\bea
M_{Z^\prime}^{}= 2 g_X  v_{\rm B-L}.
\label{MZp}
\eea
From Eq.~\eqref{eq:LYk}, it follows that after the spontaneous breaking of the $U(1)_{\rm B-L}$ and electroweak symmetries, the Majorana masses of the RHNs and the Dirac masses of the light neutrinos are generated as
\begin{equation}
    M_{N_\alpha}^{} \ = \ \frac{Y_{N_\alpha}}{\sqrt{2}} v_{\rm B-L}^{}, \, \, \, \, \,
    m_{{D}_{\alpha \beta}} \  =  \ \frac{Y_{{\nu}_{\alpha \beta}}}{\sqrt{2}} v\,.
\label{eq:mDI}
\end{equation}
which participates in the so-called seesaw mechanism~\cite{Gell-Mann:1979vob, Sawada:1979dis, Mohapatra:1980yp}. Using stationary conditions we minimize the scalar potential to obtain the mass-squared matrix for the scalar fields
\begin{align}
M^2=\begin{bmatrix}
2\lambda_H v^2 &  \lambda_{\rm mix} v  v_{\rm B-L} \\
\lambda_{\rm mix} v v_{\rm B-L}  &  2\lambda_{\Phi}v_{\rm B-L}^2 \\
\end{bmatrix}
\end{align}
for the physical states $(h_1, h_2)$ which are related to $(h, \phi)$ states following the $2\times2$ orthogonal rotation matrix $O_R$ as
\begin{align}
\begin{bmatrix}
h \\
\phi \\
\end{bmatrix}=O_R
\begin{bmatrix}
h_1 \\
h_2 \\
\end{bmatrix}=
\begin{bmatrix} \cos\theta  &  -\sin\theta \\
\sin\theta  &  \cos\theta  \\  \end{bmatrix}
\begin{bmatrix}
h_1 \\
h_2 \\
\end{bmatrix},
\label{eq:rotation}
\end{align}
where $\theta$ is the mixing angle between these two scalars. The rotation matrix satisfies $O_R M^2 O_R^T=\text{diag}\left(m_{h_1}^2,m_{h_2}^2\right)$.

Using these relations we solve for the parameters $\lambda_H$, $\lambda_\Phi$ and $\lambda_{\rm mix}$ in terms of $\theta$ and the scalar masses $m_{h_i}$ as
%%%%%%%%%%%%%%%%%%%%%%%%%%%%%%%%%%%%%%%%%%%%%%%
\begin{align}
 \lambda_{H}&=\frac{m_{h_{2}}^{2}}{4v^{2}}(1-\text{cos}~2\theta)+\frac{m_{h_{1}}^{2}}{4v^{2}}(1+\text{cos}~2\theta) \\
\lambda_{\Phi}&=\frac{m_{h_{1}}^{2}}{4v_{\rm B-L}^{2}}(1-\text{cos}~2\theta)+\frac{m_{h_{2}}^{2}}{4v_{\rm B-L}^{2}}(1+\text{cos}~2\theta)  \\
\lambda_{\rm mix}&=\text{sin}~2\theta\left(\frac{m_{h_{1}}^{2}-m_{h_{2}}^{2}}{2v v_{\rm B-L}}\right).
\label{coup}
\end{align}
where we consider that $m_{h_1}\simeq 125$ GeV is the SM like Higgs boson and $m_{h_2}$ is the BSM scalar mass and it is a free parameter.
%%%%%%%%%%%%
\begin{figure*}[htb!]
\centering
\includegraphics[scale=0.48]{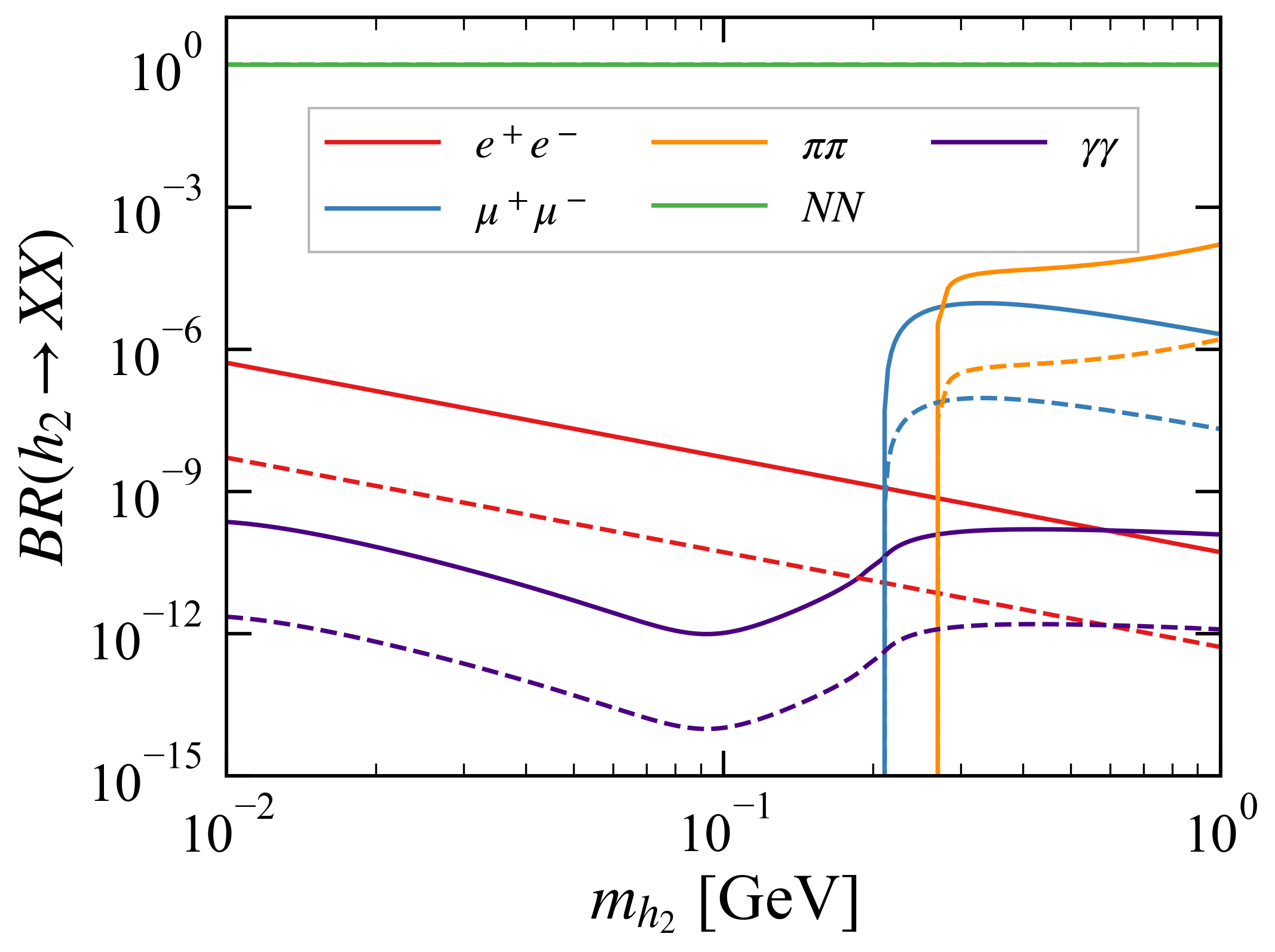}
\includegraphics[scale=0.49]{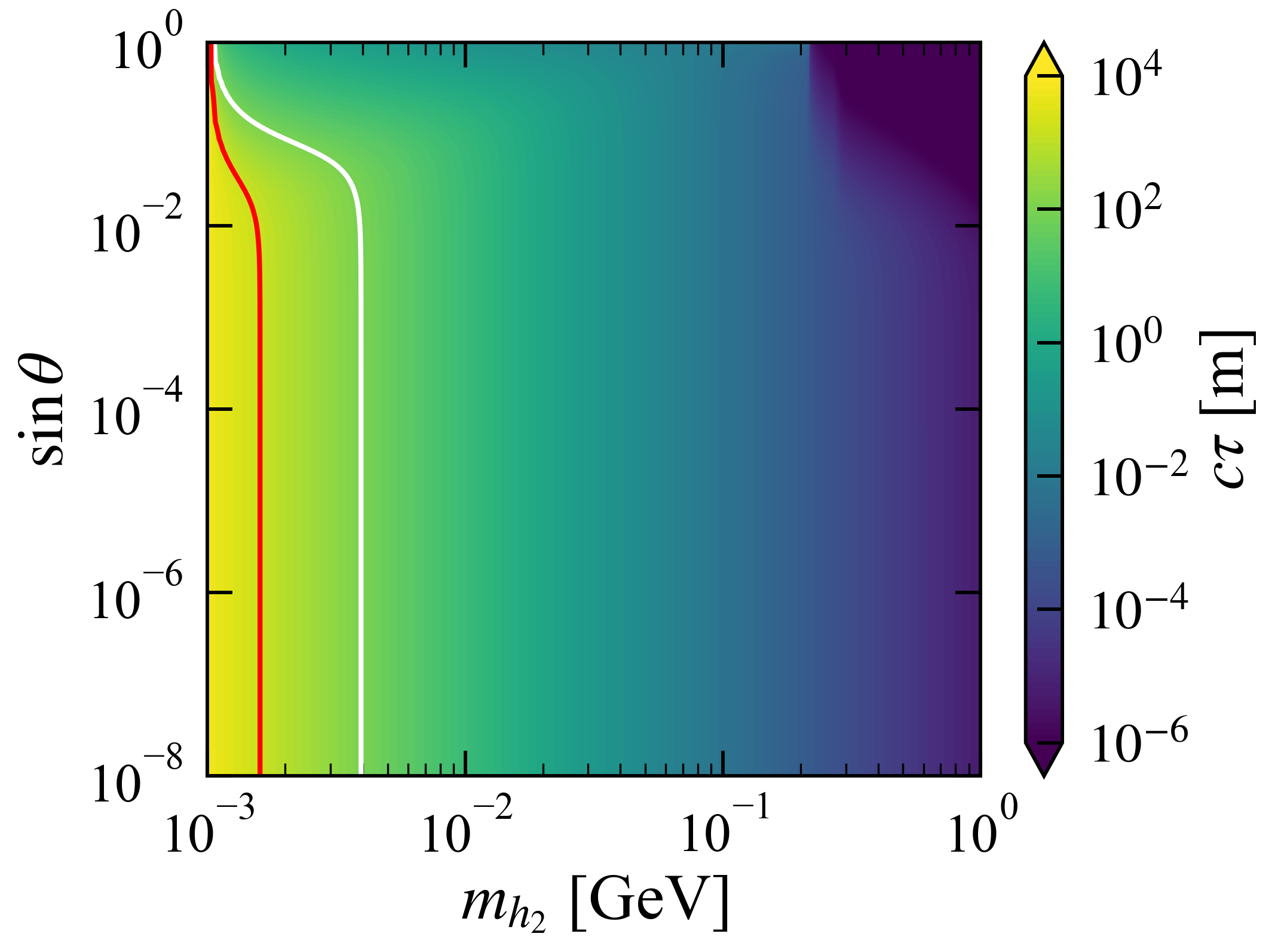}
\caption{Left panel: branching ratios of scalar $h_2$ into $ee$, $\mu\mu$, $\gamma\gamma$, $\pi\pi$~($\pi^+\pi^-+\pi^0\pi^0$) and pair of RHNs $(NN)$ as a function of scalar mass, where the solid and dashed line stands for $\sin^2\theta=10^{-8}$ and $10^{-10}$, respectively. Right panel: $h_2$ decay length in rest frame as functions of mixing $\sin\theta$ and scalar mass $m_{h_2}$.  The red~(white) contour stands for the decay length $L_{h_2}$ = 1.5 km~(100 m) relevant for FPF~(SHiP) detector. For both the panels we fixed other relevant parameters as follows: $M_{Z^\prime} =100$ GeV,  $g_X=0.01$ and  $M_N=m_{h_2}/4$.
}
\label{branching-scalar}
\end{figure*}
%%%%%%%%%%%%%%%%%%%%%%%%%%%%%
\begin{figure*}[htb!]
\centering
\includegraphics[scale=0.48]{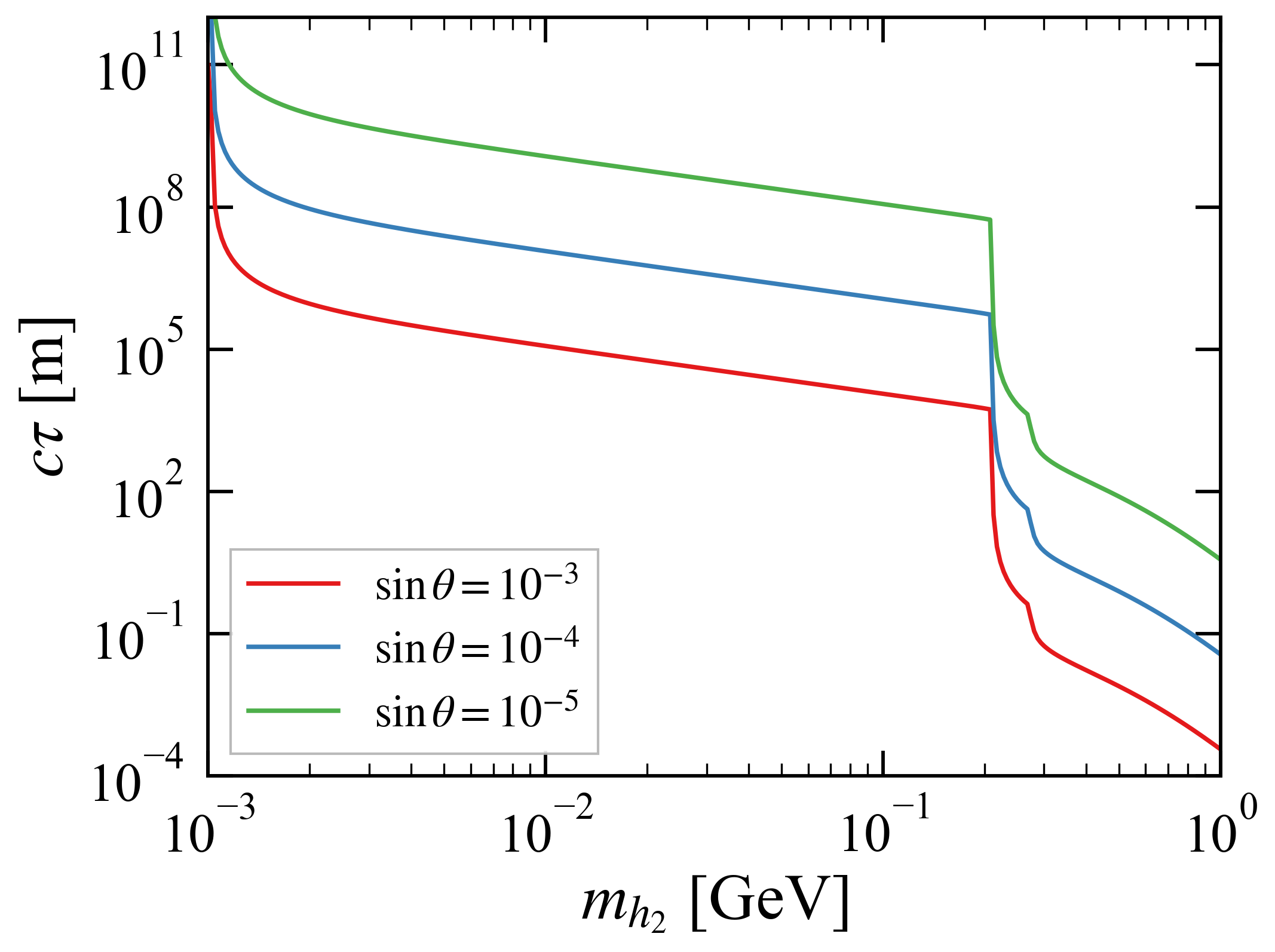}~~
\includegraphics[scale=0.48]{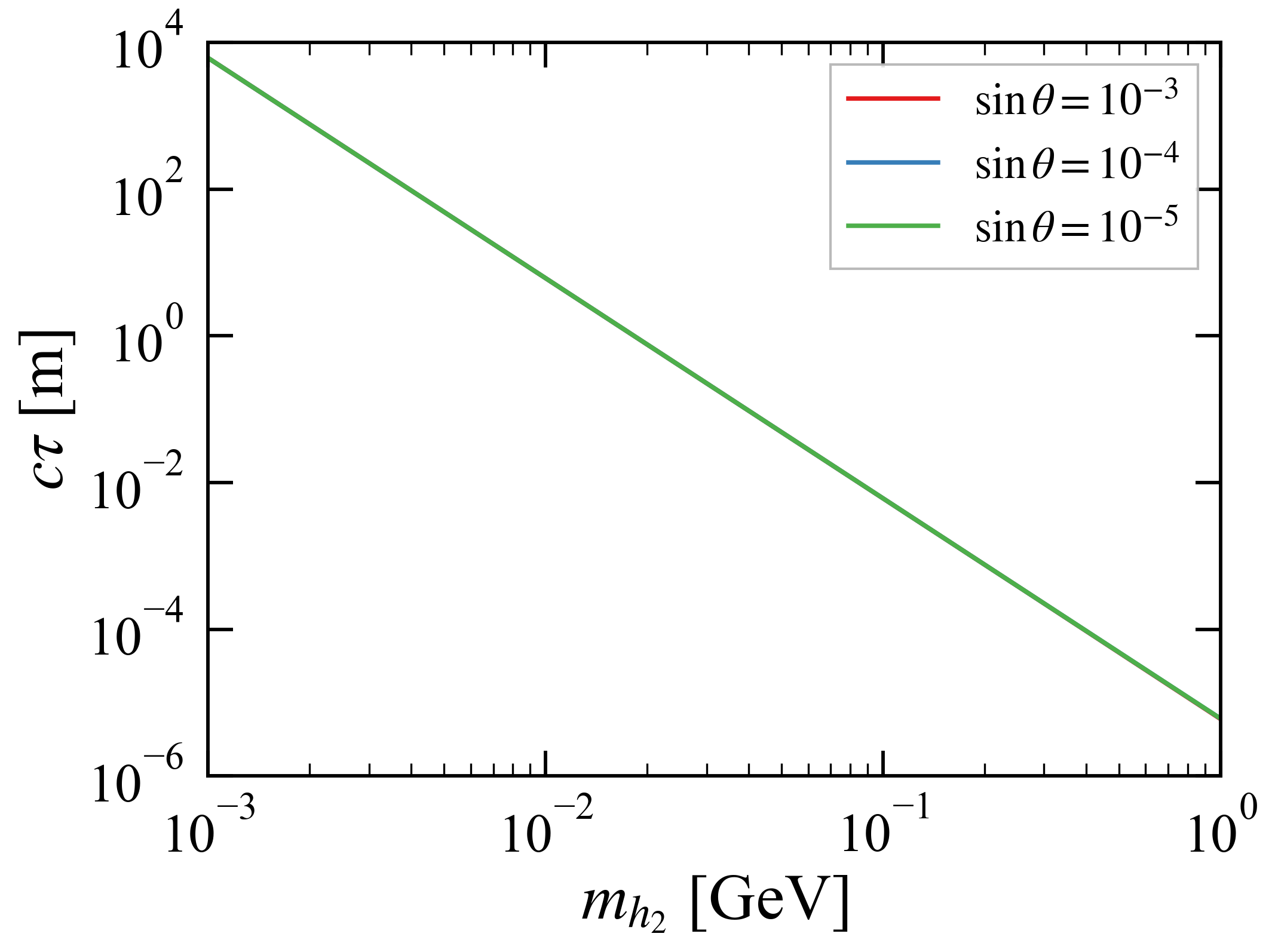}
\caption{Decay length of $h_2$ as a function of its mass without (left panel) and with (right panel) RHN decay channels, for three representative values of the scalar mixing angle, $\sin\theta=10^{-3}$ (red), $10^{-4}$ (blue), and $10^{-5}$ (green). In the right panel, the three curves coincide.}
\label{h2length}
\end{figure*}
%%%%%%%%%%%%%%%%%
\begin{figure*}[htb!]
\centering
\includegraphics[scale=0.45]{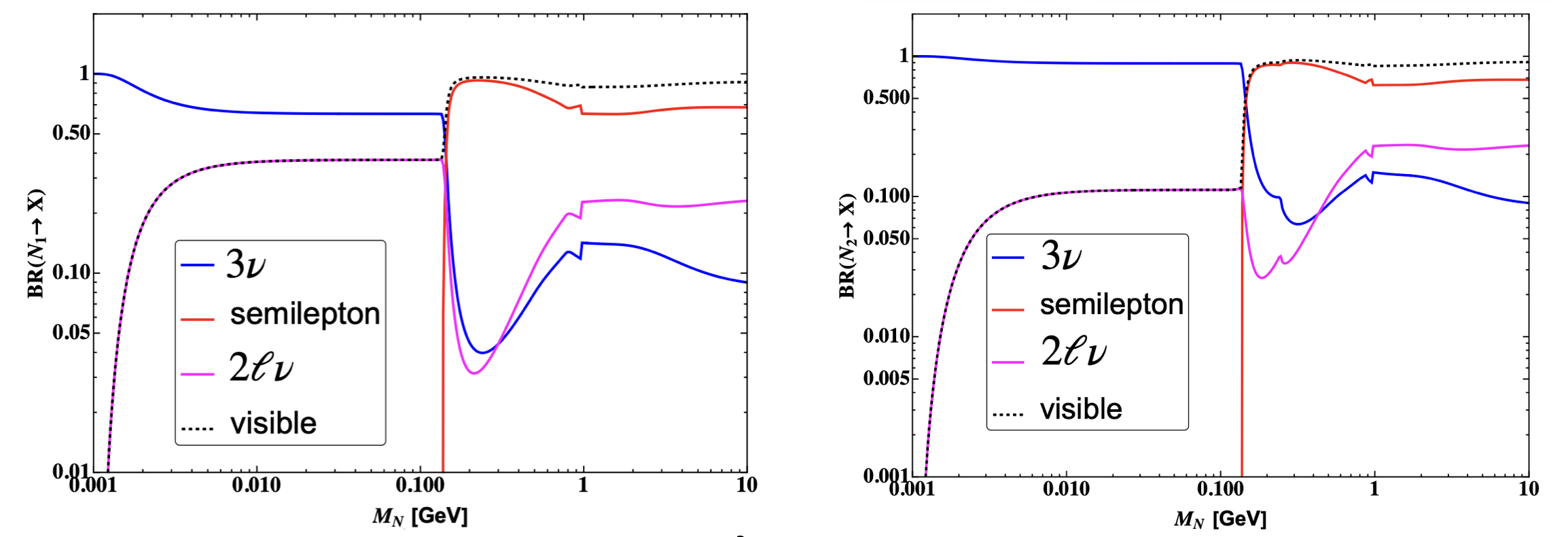}
\caption{Branching ratio of RHNs in different modes. The dashed black line in each of the panel stands for the branching ratio of the RHN to the visible mode.}
\label{RHN-decay}
\end{figure*}
%%%%%%%%%%%%%%%%%

Under $U(1)_{\rm B-L}$ scenario the neutral BSM gauge boson $Z^\prime$ interacts with the SM fermions $(f_{L(R)})$ following
\bea
\mathcal{L} = -g_X (\overline{f_L}\gamma^\mu q_{f_{L}^{}}^{}  f_L+ \overline{f_R}\gamma^\mu q_{f_{R}^{}}^{}  f_R) Z_\mu^\prime.
\label{Lag1}
\eea
where $q_{f_{L(R)}^{}}^{}$ is the $U(1)_{\rm B-L}$ charge of the left (right) handed fermions. The B$-$L scenario is a vector like where left and right handed fermions have same $U(1)_{\rm B-L}$ charge. The partial decay widths of $Z^\prime$ into  charged fermions can be written as
\begin{align}
\Gamma(Z^\prime \to \bar{f} f) &= N_C \frac{M_{Z^\prime} g_{X}^2}{24 \pi} \left[ \left( q_{f_L}^2 + q_{f_R}^2 \right) \left( 1 - \frac{m_f^2}{M_{Z^\prime}^2} \right) \right. \nonumber \\
&\quad \left. + 6 q_{f_L} q_{f_R} \frac{m_f^2}{M_{Z^\prime}^2} \right] \sqrt{1 - \frac{4 m_f^2}{M_{Z^\prime}^2}},
\label{eq:width-ll}
\end{align}
where $N_C$ is the color factor for SM fermions and $m_f$ is the mass of the corresponding fermions. The partial decay width of $Z^\prime$ into a pair of single generation light neutrinos can be derived from Eq.~(\ref{eq:width-ll}) neglecting the tiny neutrino mass and putting $q_{f_R}=0$. %can be written as
%\begin{align}
%\label{eq:width-nunu}
%    \Gamma(Z^\prime \to \nu \nu)
%    =  \frac{M_{Z^\prime}^{} g_{X}^2}{24 \pi} q_{f_L^{}}^2~,
%\end{align}
%where $q_{f_L^{}}^{}$ is the general $U(1)$ charge of the SM lepton doublet $\ell_L$.
In general $U(1)$ extended SM scenarios, $Z^\prime$ interacts with heavy neutrinos following
\bea
\mathcal{L}_N= -\frac{1}{2}g_X q_{N_R} \overline{N} \gamma_\mu \gamma_5 N Z_{\mu}^\prime.
\label{neut}
\eea
where $q_{N_R}(=-1)$ is the B$-$L charge of the RHNs. The partial decay width of $Z^\prime$ into a pair of single generation heavy neutrino is
\begin{align}
\label{eq:width-NN}
    \Gamma(Z^\prime \to N^\alpha N^\alpha)
    = \frac{ g_{X}^2 M_{Z^\prime}^{}}  {24 \pi} \left( 1 - \frac{4 M_{N_\alpha}^2}{M_{Z^\prime}^2} \right)^{\frac{3}{2}}~,
\end{align}
where $M_{N_\alpha}$ being the heavy neutrino mass.
%%%%%%%%%%%%%%%%%%%%%%%%%%%%%%%%%%%%%%%%%%%%%%%%%%%%%%%%%%%%%%%%%%%%%%%%%%%%%%%%%%%%%%%

In this analysis we consider $m_{h_2}\ll m_{h_1} \ll M_{Z^\prime}$ and $M_{N_\alpha} < m_{h_2}/2$ so that the decay mode $h_2 \to N N$ could be kinematically allowed. Owing to its mixing with the SM Higgs boson, the scalar $h_2$ can also decay into SM leptons, pion pairs, and two photons, where the diphoton channel proceeds via charged fermion and $W$ boson loops. Partial decay widths of $h_2$ into different modes can be given by
\bea
\Gamma(h_2 \to N_\alpha N_\alpha)&=& \frac{Y_{N_\alpha}^2 m_{h_2} \cos^2\theta}{32 \pi}  \Big(1-4\frac{m_{N_\alpha}^2}{m_{h_2}^2}\Big)^{3/2}, \\
\Gamma(h_2 \to \SM~\SM)&=& \sin^2\theta \times \tilde{\Gamma}(h_2 \to \SM~\SM).
\eea
where the analytical expressions of each of these decay widths into different SM modes can be found in the Appendix~\ref{sec:decays1}. As a result of the gauge coupling of $h_2$ to $Z'$ bosons, if kinematically allowed, we have also the decay channel $h_2\to Z' Z'$. It is worth emphasizing that all decay channels of $h_2$ into SM particles scale universally with the mixing angle $\sin\theta$, whereas the $h_2 \to Z'Z'$ decay is controlled by the $U(1)_{B-L}$ gauge coupling $g_{X}$. Therefore, when $g_{X}\to 0$ or the decay $h_2 \to Z'Z'$ is kinematically forbidden, the scalar $h_2$ decays only into SM particles and RHNs. In this work, we focus on this scenario.

The branching ratios of $h_2$ as functions of its mass, $m_{h_2}$, are displayed in the left panel of Fig.~\ref{branching-scalar} for two representative benchmark values of the scalar mixing angle, $\sin^2\theta=10^{-8}$ (solid curves) and $\sin^2\theta=10^{-10}$ (dashed curves) where we considered RHNs mass as $M_N=m_{h_2}/4$ and fixed the $Z'$ mass at 100 GeV. We find that as long as the scalar mixing is small and $h_2\to Z'Z'$ mode is forbidden, the $h_2\to NN$ always dominates for our interested mass range of $h_2$. Finally, using the total decay width of the $h_2$, we compute its proper decay length by assuming that the $h_2$ decays in its rest frame. In the right panel of Fig.~\ref{branching-scalar}, we show the decay length of $h_2$ as a function of mass and scalar mixing. We find that for $m_{h_2}<10^{-2}$ GeV, the scalar $h_2$ is typically long-lived, largely independent of the scalar mixing angle. Such long-lived scalars can be probed at FPF@FCC (SHiP) when their decay length is approximately $1.5$--$2.0$ km ($100$ m), as indicated by the red (white) contour in Fig.~\ref{branching-scalar}. In contrast, for $m_{h_2}>10^{-2}$ GeV, $h_2$ is generally short-lived, again almost independent of the scalar mixing angle. This behavior arises because the decay channel $h_2\to NN$ dominates over most of the parameter space and its partial decay width is essentially insensitive to the scalar mixing angle.

To further clarify this point, we include Fig.~\ref{h2length}, which shows the decay length of $h_2$ for different values of the scalar mixing angle, without (left panel) and with (right panel) the inclusion of the $h_2\to NN$ decay mode. The left panel illustrates that, in the absence of the RHN channel, $h_2$ decays only into SM particles through the scalar mixing $\sin\theta$. Consequently, its decay length increases as the mixing decreases, allowing $h_2$ to be either short-lived or long-lived depending on the value of $\sin\theta$, even for relatively large scalar masses. In contrast, once the $h_2\to NN$ channel is kinematically accessible, it dominates the total decay width, making the $h_2$ decay length nearly independent of the scalar mixing angle. As a result, $h_2$ is generally short-lived except for very light masses or very small Yukawa $Y_N$. Therefore, unlike conventional LLP scenarios in which the BSM scalar itself is long-lived and decays into SM particles inside the detector, our scenario features a promptly decaying scalar that predominantly produces a pair of RHNs. These RHNs can be long-lived and travel macroscopic distances before decaying into visible final states inside the detector, providing the characteristic LLP signature studied in this work.

Being SM-singlet RHNs do not interact with the SM sector directly,
however, from the seesaw mechanism, we can write the light neutrino flavor eigenstate $(\nu)$ in terms of the mass eigenstates of the light $(\nu_m)$ and heavy $(N_m)$ neutrinos as $\nu \simeq  \nu_m  + V_{\ell N} N_m$
where $V_{\ell N}(=m_D/m_N)$ is the mixing between the light and heavy mass eigenstates, and we assume $|V_{\ell N}| \ll 1$. This will introduce new charged current (CC) and neutral current (NC) interactions as follows
\begin{align}
\mathcal{L}_{\rm CC}^N & \supset
 -\frac{g}{\sqrt{2}} W_{\mu}
  \bar{\ell} \gamma^{\mu} P_L V_{\ell N} N_m  + {\rm H.c.},
\label{CC}\\
\mathcal{L}_{\rm NC}&\supset
 -\frac{g}{2 \cos\theta_{\rm W}}  Z_{\mu}
\Big[
  \overline{N_m} \gamma^{\mu} P_L  |V_{\ell N}|^2 N_m \nonumber \\
&+ \Big\{
  \overline{\nu_m} \gamma^{\mu} P_L V_{\ell N}  N_m
  + {\rm H.c.}\Big\} \Big] ,
\label{NC}
\end{align}
where $\ell$ denotes the three generations of the charged leptons in the vector form, and
$P_L =\frac{1}{2} (1- \gamma_5)$ is the projection operator. From Eqs.~\eqref{CC} and \eqref{NC}, the RHNs can decay into the SM particles through the mixing. For our interested mass range, the RHNs can decay into purely leptonic modes such as $N \to \nu \ell^+ \ell^-$, $N \to 3\nu$ or semileptonic modes as $N \to \nu \mathscr{H}^{0}$, $N \to \ell^\pm \mathscr{H}^{\mp}$ with neutral (charged) meson ($\mathscr{H}^{0(\pm)}$), respectively.
The corresponding partial decay widths of all possible modes are calculated in \cite{Bloom:1970xb,Bloom:1971ye,Braaten:1991qm,Cvetic:2001sn,Dib:2000wm,Ivanov:2004ch,Gribanov:2001vv,Cvetic:2010rw,Helo:2010cw,A:2025ygb}.
%%%%%%%%%%%%%%%%%%%%%%%%%
\begin{figure*}[htb!]
\centering
\includegraphics[scale=0.186]{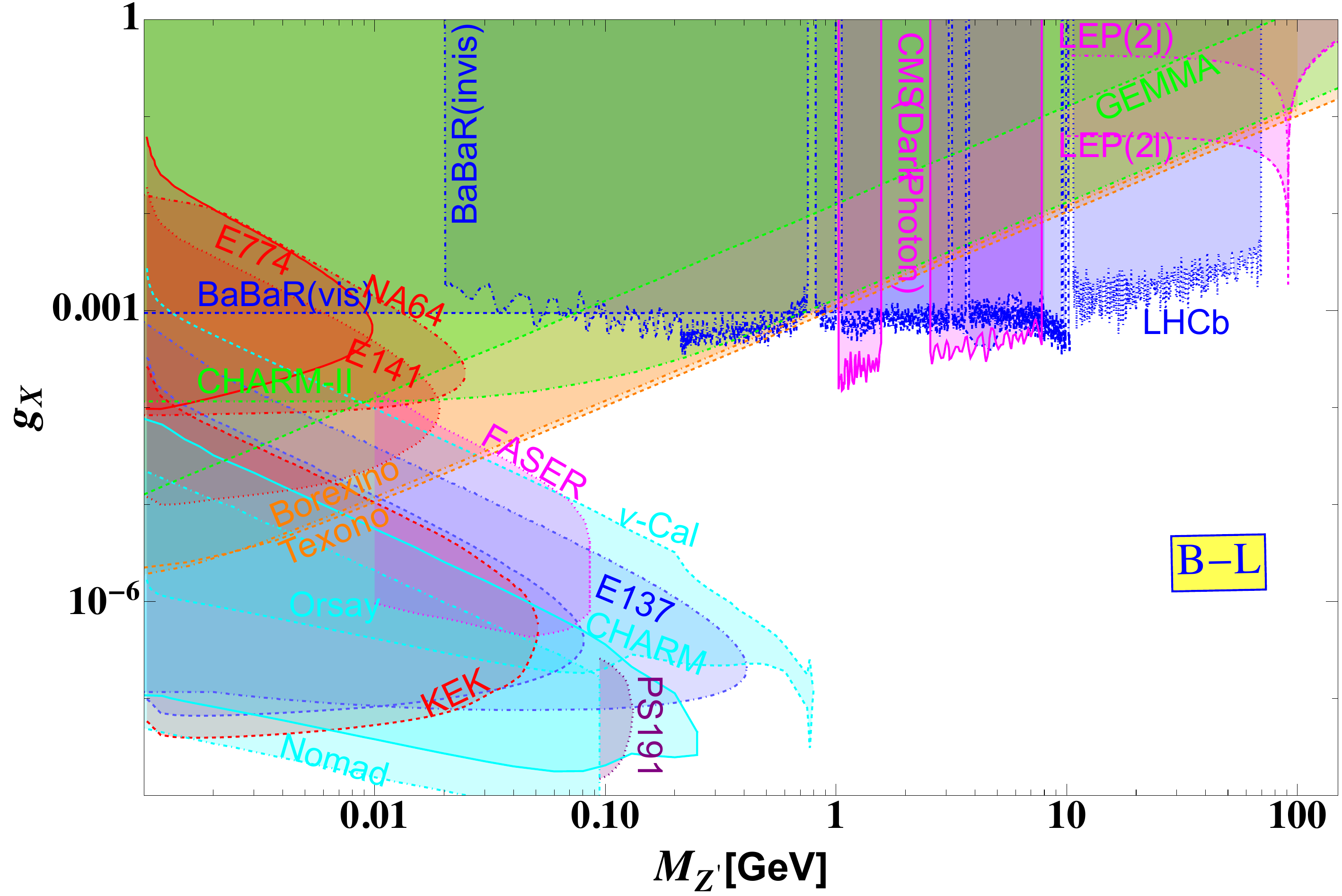}
\includegraphics[scale=0.185]{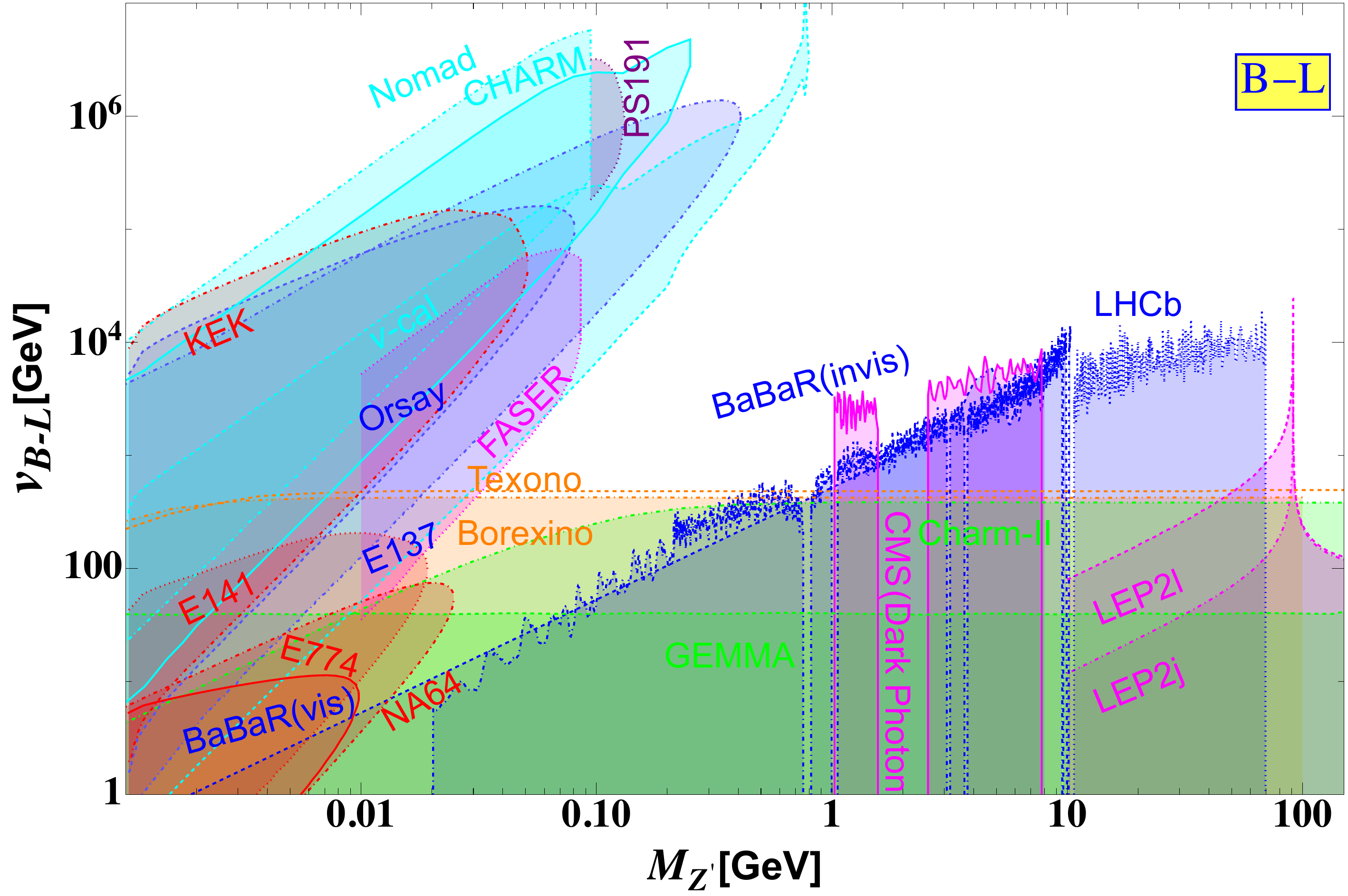}
\includegraphics[scale=0.185]{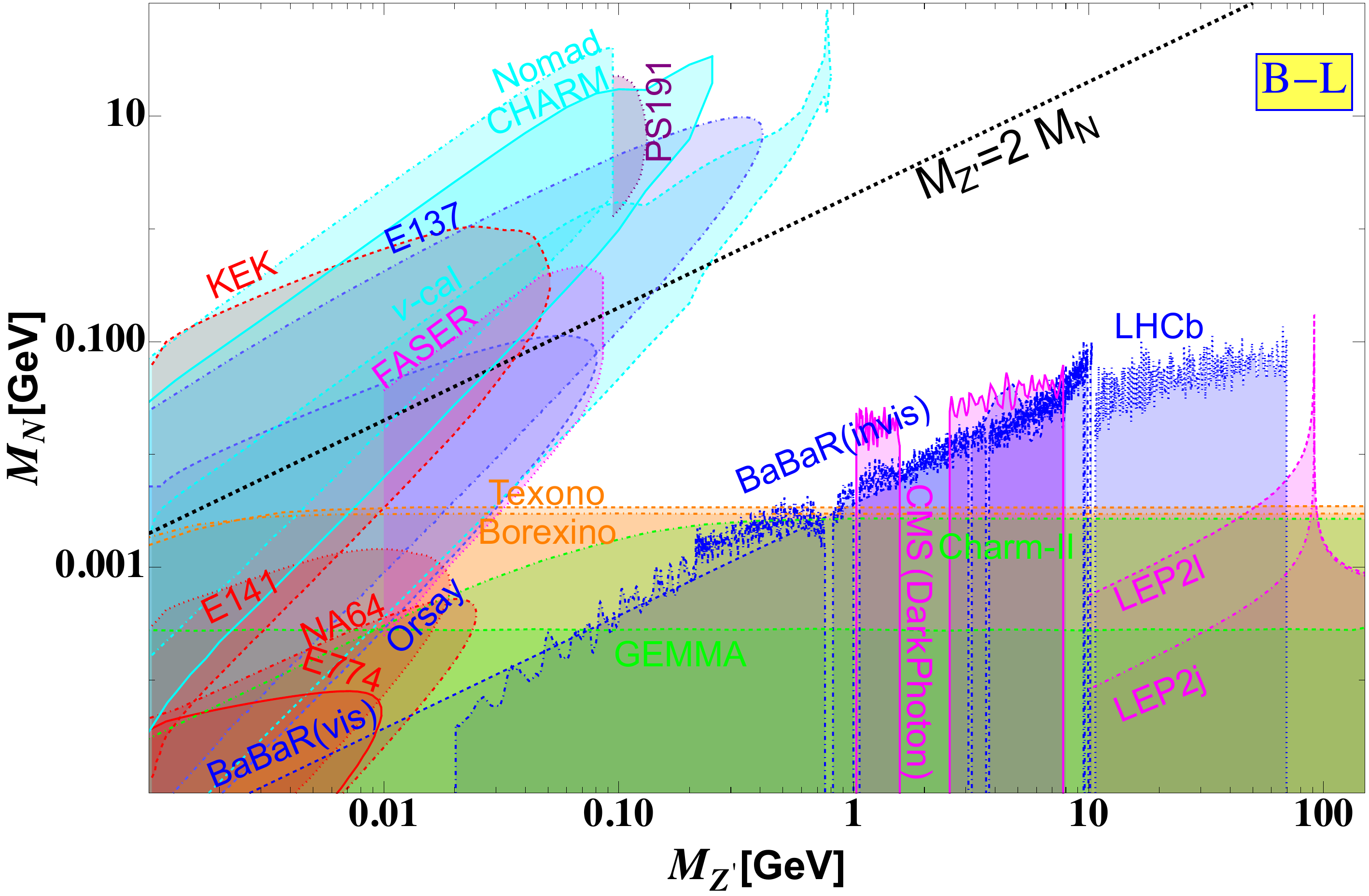}
\includegraphics[scale=0.172]{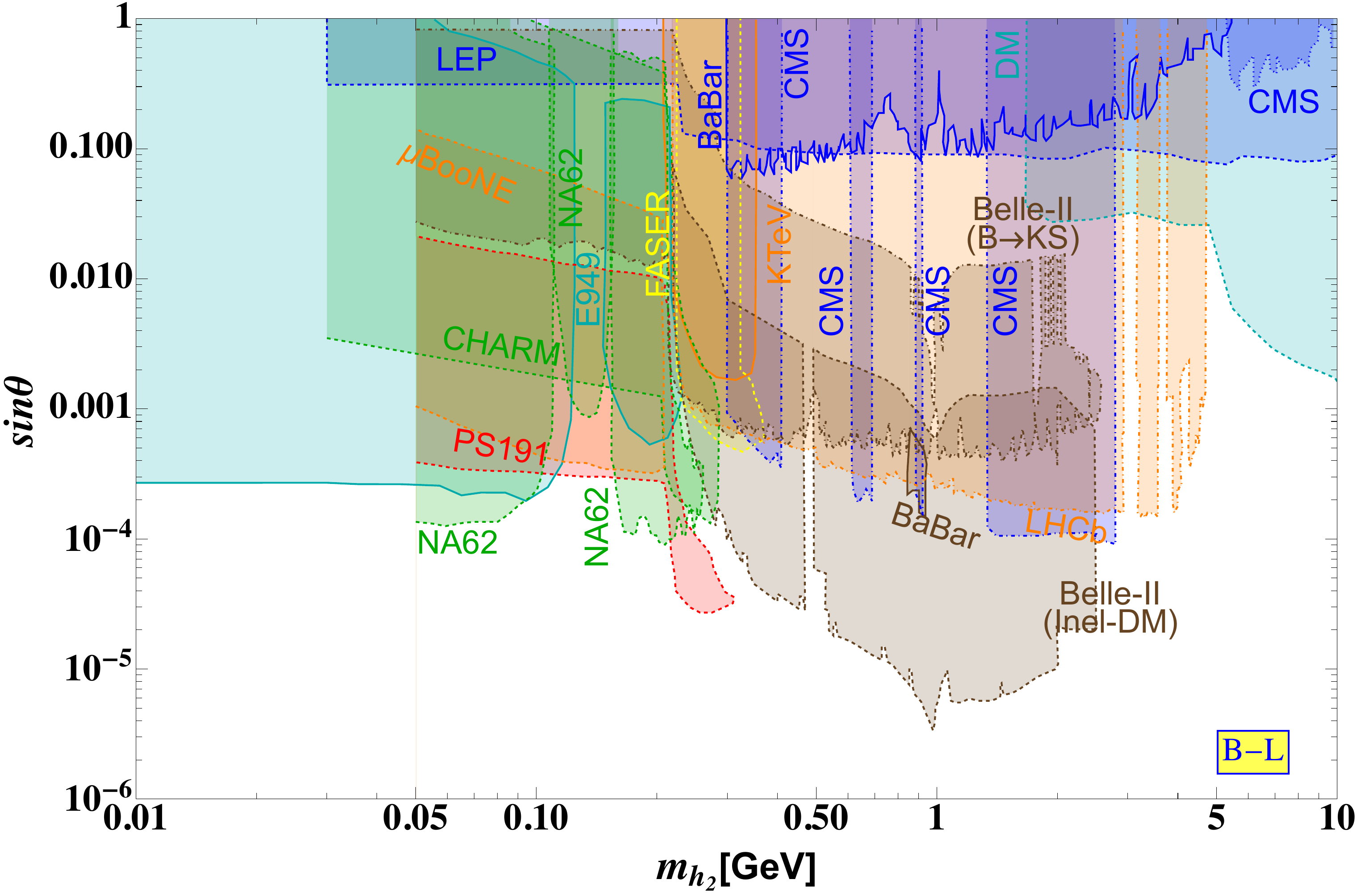}
\caption{Existing bounds from various experiments on model parameters: (i) $g_X-M_{Z^\prime}$ (top left), (ii) $v_{\rm B-L}-M_{Z^\prime}$ (top right), (iii) $M_N-M_{Z^\prime}$ (bottom left) and (iv) $\sin\theta-m_{h_2}$ considering $h_2 \to$ SM SM modes (bottom right), respectively for the B$-$L scenario.}
\label{params}
\end{figure*}
%%%%%%%%%%%%%%%%%%%%%%%%%%%%

Fig.~\ref{RHN-decay} displays the branching ratios of the RHNs, assuming that the first and second-generation RHNs couple predominantly to electrons and muons, respectively. Since each heavy neutrino is taken to mix with only one charged-lepton flavor, the corresponding branching ratios are independent of the magnitude of the light--heavy neutrino mixing. Below the pion-mass threshold, the invisible three-neutrino decay mode dominates, whereas semileptonic channels become increasingly important once the pion channel opens. The total branching fraction of $N_{i=1,2}$ into visible final states is shown by the black dashed curve. This quantity is particularly relevant for estimating the expected signal yield, since experimental searches are primarily sensitive to visible decay products rather than the invisible $3\nu$ final state. We observe that for $m_{N_i}$ above the light-hadron threshold (e.g., $\pi$ and $\eta$), the branching fraction of $N_i$ into visible final states exceeds $90\%$.

For sufficiently small light-heavy neutrino mixing, the RHN becomes a long-lived particle that can travel macroscopic distances before decaying. The FPF@FCC (SHiP) experiments are particularly sensitive to such long-lived heavy neutrinos with decay lengths of approximately 1.5--2.0 km (100 m), depending on the detector configuration.

%%%%%%%%%%%%%%%%%%%%%%%%%%%%%
\section{Existing limits}
%%%%%%%%%%%%%%%%%%%%%%%%%%
We summarize the current experimental constraints on the model parameters in Fig.~\ref{params}, where the excluded regions are indicated by the shaded areas. The upper-left panel shows the existing limits in the $g_X$--$M_{Z^\prime}$ plane. Constraints from electron beam-dump experiments, including Orsay \cite{Davier:1989wz}, KEK \cite{Beer:1986qr}, E141 \cite{Riordan:1987aw}, E137 \cite{Bjorken:1988as}, and E774 \cite{Bross:1989mp}, are presented together with bounds from proton beam-dump experiments such as $\nu$-Cal \cite{Blumlein:2011mv,Blumlein:2013cua}, LSND \cite{LSND:1997vqj}, PS191 \cite{Bernardi:1985ny}, NOMAD \cite{NOMAD:2001eyx}, and CHARM \cite{CHARM:1985anb}. We also include constraints from neutrino scattering experiments. The neutrino--electron scattering limits are derived from TEXONO \cite{TEXONO:2009knm}, BOREXINO \cite{Borexino:2008gab,Bellini:2011rx}, GEMMA \cite{Beda:2010hk}, and CHARM-II \cite{CHARM-II:1994dzw}, while neutrino--nucleon scattering constraints are taken from COHERENT \cite{COHERENT:2020ybo}. Finally, we show bounds from dark photon searches performed at LHCb \cite{LHCb:2019vmc}, BaBar \cite{BaBar:2014zli,BaBar:2017tiz}, and CMS \cite{CMS:2023slr}.

We also derive constraints on the $g_X$--$M_{Z^\prime}$ plane from precision measurements at the $Z$ pole. We calculate the cross section for the process $e^-e^+ \to f\bar{f}$, including the contribution from the $Z^\prime$ boson in the $B-L$ scenario, and compare it with the LEP measurements reported in Ref.~\cite{ALEPH:2013dgf}. We first consider the hadronic final state, $e^-e^+ \to q\bar{q}$, for which the measured cross section is $\sigma = 41.544 \pm 0.037$ nb. Requiring the $Z^\prime$ contribution to remain consistent with the measured value, we derive the corresponding $90\%$ C.L. upper limits, shown by the magenta dot-dashed curve in the upper-left panel of Fig.~\ref{params}. We next consider the dilepton process, $e^-e^+ \to \ell^-\ell^+$, using the precisely measured ratio $R_\ell=\Gamma_{\rm had}/\Gamma_\ell = 20.768 \pm 0.025$ to obtain complementary constraints. The resulting $90\%$ C.L. limits are shown by the magenta dashed curve in the same panel. Owing to the high precision of the leptonic observables, the dilepton channel provides the strongest constraints in the vicinity of the $Z$ pole.

The expression for the $Z^\prime$ mass in Eq.~\eqref{MZp} is derived under the assumption that $v_{\rm B-L} \gg v$. Using this relation, we translate the existing constraints in the $g_X$--$M_{Z^\prime}$ plane into the $V_{\rm B-L}$--$M_{Z^\prime}$ plane. The resulting exclusion limits are shown in the upper-right panel of Fig.~\ref{params}, where the excluded regions are indicated by the shaded areas. Combining Eqs.~\eqref{MZp} and \eqref{eq:mDI}, the heavy neutrino mass can be expressed as
\bea
M_{N_\alpha}= \frac{Y_{N_\alpha}}{2\sqrt{2}} \frac{M_{Z^\prime}}{g_X}.
\eea
Assuming a representative Yukawa coupling $Y_{N_\alpha}=10^{-5}$, we further map the experimental constraints onto the $M_N$--$M_{Z^\prime}$ plane, as shown in the lower-left panel of Fig.~\ref{params}. These limits are obtained using the experimentally allowed values of the ratio $M_{Z^\prime}/g_X$ derived from the constraints in the upper-left panel. The black dashed line corresponds to the kinematic threshold $M_{Z^\prime}=2M_N$, below which the decay channel $Z^\prime\to NN$ becomes kinematically accessible.

We summarize the existing constraints on the scalar mixing angle in the lower-right panel of Fig.~\ref{params}. These limits are obtained in the framework of the SM extended by a singlet scalar that mixes with the SM Higgs boson and subsequently decays into SM particles. The excluded regions, indicated by the shaded areas, include bounds from direct searches at CHARM \cite{CHARM:1985anb}, LHCb \cite{LHCb:2015nkv,LHCb:2016awg,LHCb:2018cjc,Gorkavenko:2023nbk}, L3 \cite{Ferber:2023iso,L3:1996ome}, NA62 \cite{NA62:2020pwi,NA62:2021zjw}, CMS \cite{CMS:2012fgd,CMS:2021sch,Winkler:2018qyg}, PS191 \cite{Gorbunov:2021ccu,BNL-E949:2009dza,Gorbunov:2023lga}, BaBar \cite{BaBar:2015jvu,Winkler:2018qyg}, KTeV \cite{KTEV:2000ngj,KOTO:2020prk}, LEP \cite{L3:1996ome}, E949 \cite{BNL-E949:2009dza}, FASER \cite{Feng:2017vli}, Belle II \cite{Belle-II:2025bhd}, and MicroBooNE \cite{MicroBooNE:2021sov,MicroBooNE:2022ctm}. For completeness, we also show the constraints from dark scalar searches \cite{delValle:2025zvo} and dark matter direct detection experiments \cite{CRESST:2017cdd,XENON:2018voc,DarkSide:2018bpj,Winkler:2018qyg}. Although our minimal setup does not include a dark matter candidate, these bounds provide useful reference points. An example of a dark matter realization within a similar framework can be found in Ref.~\cite{Barman:2024lxy}.

%%%%%%%%%%%%%%%%%%%%
\section{Results and discussion}
\label{secIII}
%%%%%%%%%%%%%%%%%%%%%%%%%%%%%%%%%%%%%%%%%%%%%%%%%%%%%%%
\begin{center}
      \begin{table*}[t]
    \centering
    \begin{tabular}{|c|c|l|l|l|l|l|}\hline
 Experiment& $L_1$& $L_2$& Surface
(front)& Surface
 (back)&$\mathcal{L}$/(N$_\text{POT}/\sigma_\text{total}$) &$F (X, Y)$\\\hline
 FPF1@FCC& 1.5 km& 50 m& circular:
5 m $\times$ 5 m& circular:
5 m $\times$ 5 m& 30 ab$^{-1}$ &$\Theta(\sqrt{X^2 + Y^2} - 2.5)$\\\hline
         FPF2@FCC&  1.5 km& 400 m& circular:
20 m $\times$ 20 m& circular:
20 m $\times$ 20 m& 30 ab$^{-1}$ &$\Theta(\sqrt{X^2 + Y^2} - 10)$\\\hline
         SHiP&
     70 m& 50 m& rectangular:
5 m $\times$ 10 m& rectangular:
6 m $\times$ 12 m& $\frac{6 \times 10^{20}}{1.085 \times 10^{18}} \approx 553$ ab$^{-1}$ &$\Theta(X - 2.5)\Theta(Y - 5)$\\ \hline\end{tabular}
\caption{Detector information involving decay volume (length, cross section), luminosity  for FPF and SHiP experiments.}
\label{tab:placeholder}
\end{table*}
\end{center}
%%%%%%%%%%%%%%%%%%%%%%%%%%%%%%%%%%%%%%%%%%%%%%%%%%%%%%%%%%%%%%%%%%%%%%%%%%%%%%%%%%%
\begin{figure*}[htb!]
\centering
\includegraphics[scale=0.173]{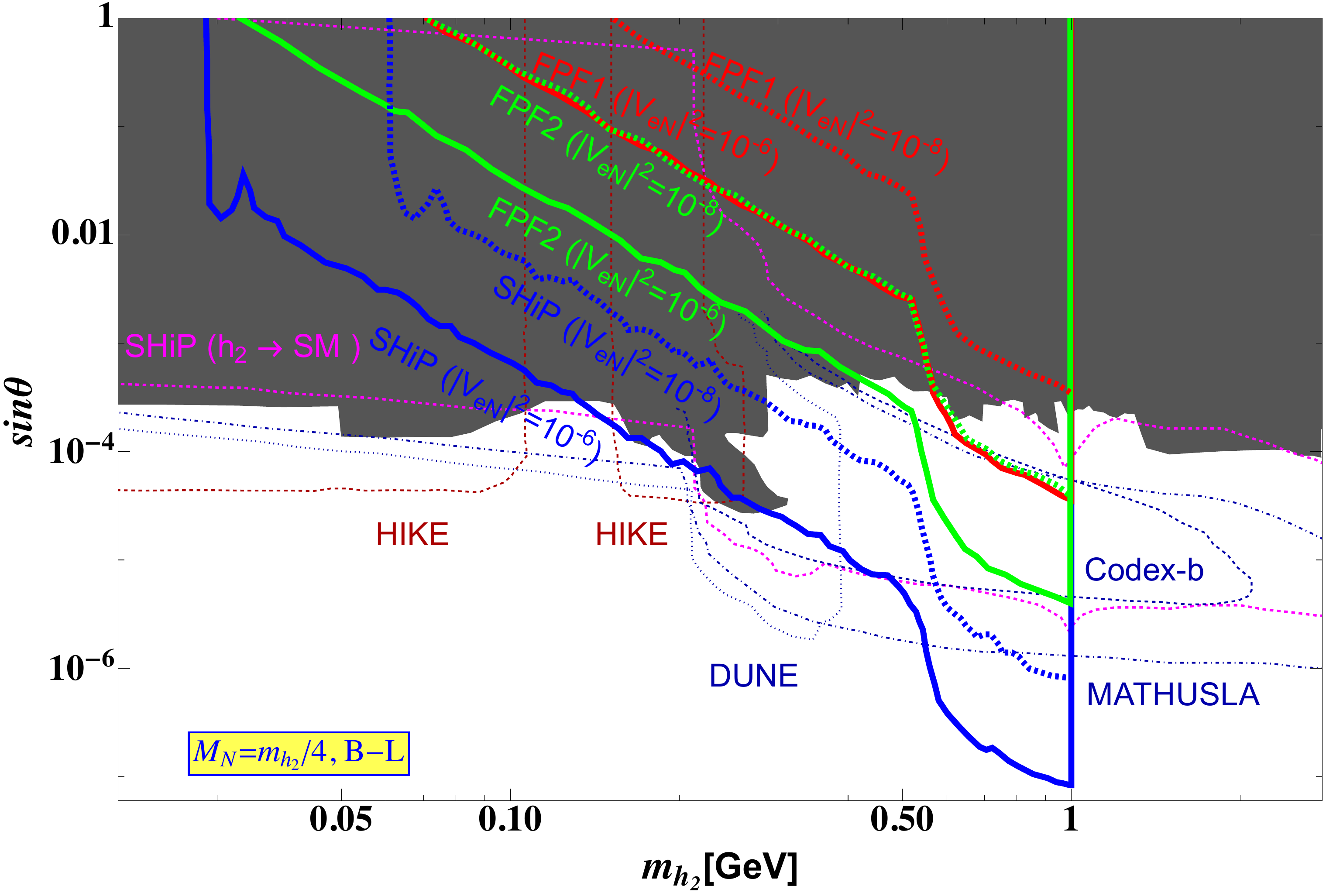}
\includegraphics[scale=0.173]{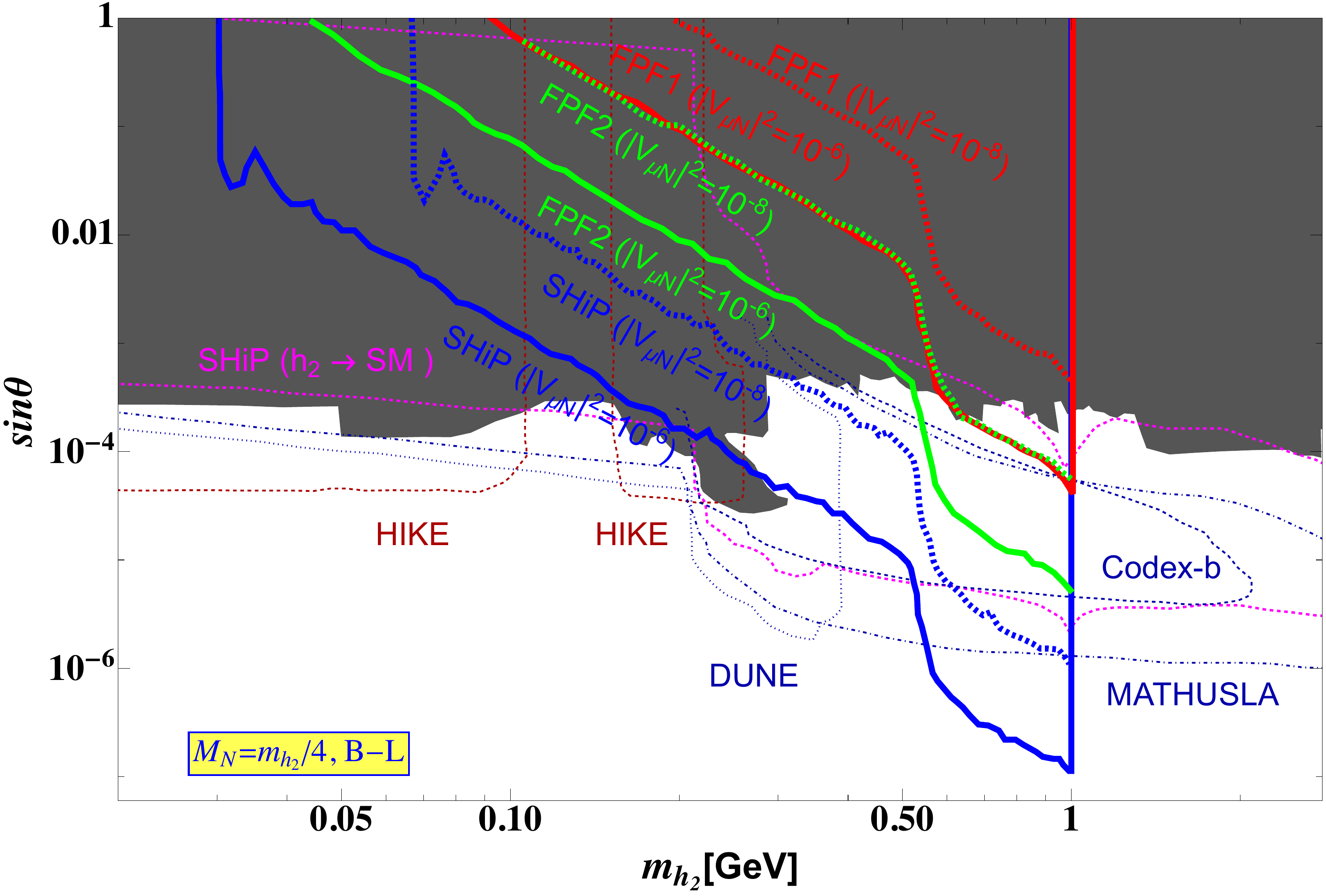}
\caption{Prospective limits on $\sin\theta$ in the left (right) panel as a function of $m_{h_2}$ for long-lived heavy neutrinos produced via the decay of a short-lived scalar and decay dominantly into electrons (muons), assuming $|V_{e(\mu) N}|^2 = 10^{-6}$ (dot-dashed) and $10^{-8}$ (dashed) for the B$-$L. The red, green and blue curves correspond to FPF1, FPF2 and SHiP experiments considering $M_{N} = m_{h_2}/4$. Dark gray shaded region is excluded by existing experimental constraints. We compare our results with prospective bounds from HIKE\cite{HIKE:2022qra}, DUNE \cite{Berryman:2019dme}, Codex-b \cite{CODEX-b:2019jve}, MATHUSLA \cite{Bertuzzo:2020rzo,Curtin:2018mvb} and  SHiP \cite{SHiP:2015vad} considering $h_2 \to$ SM SM mode.}
\label{mix1}
\end{figure*}
%%%%%%%%%%%%%%%%%%%%%%%%%%%%%%%%%%%%%%%%%%%%%%%%%%%%%%
\begin{figure*}[htb!]
\centering
\includegraphics[scale=0.37]{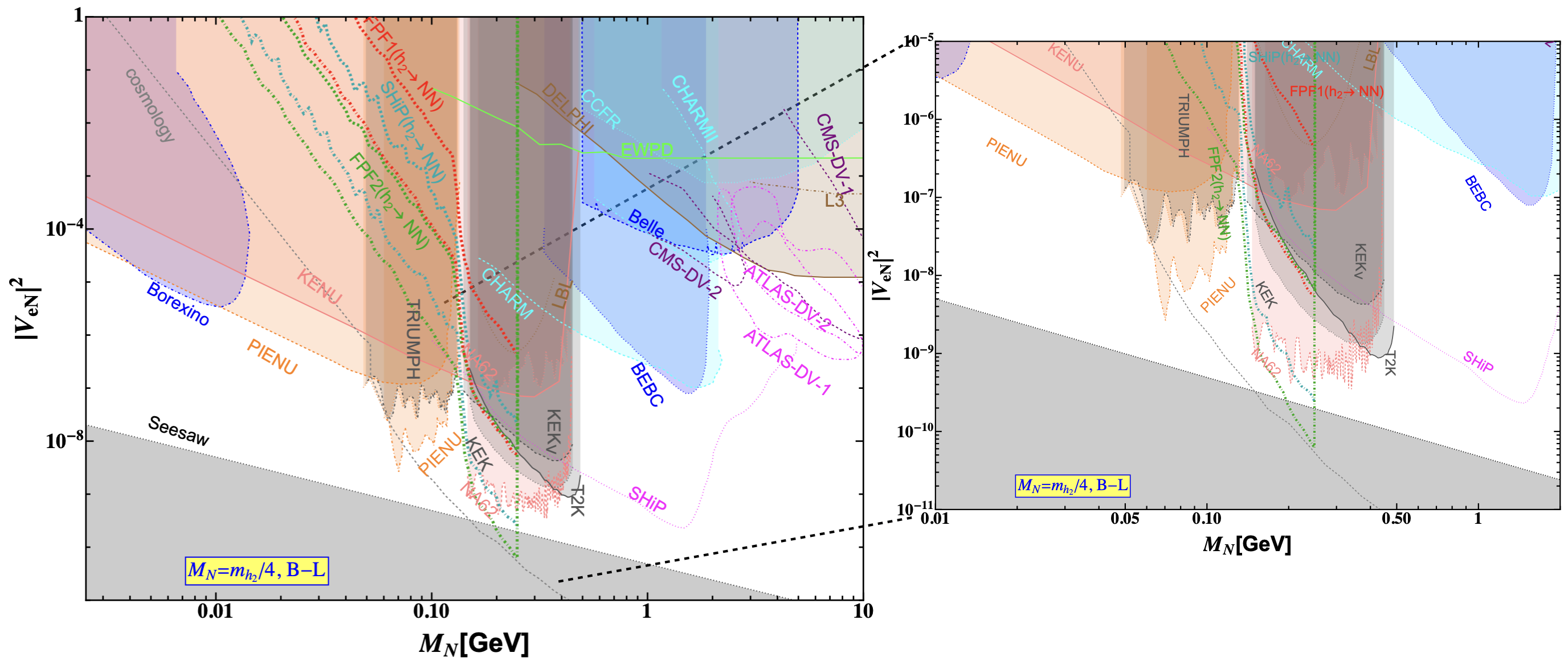}
\includegraphics[scale=0.37]{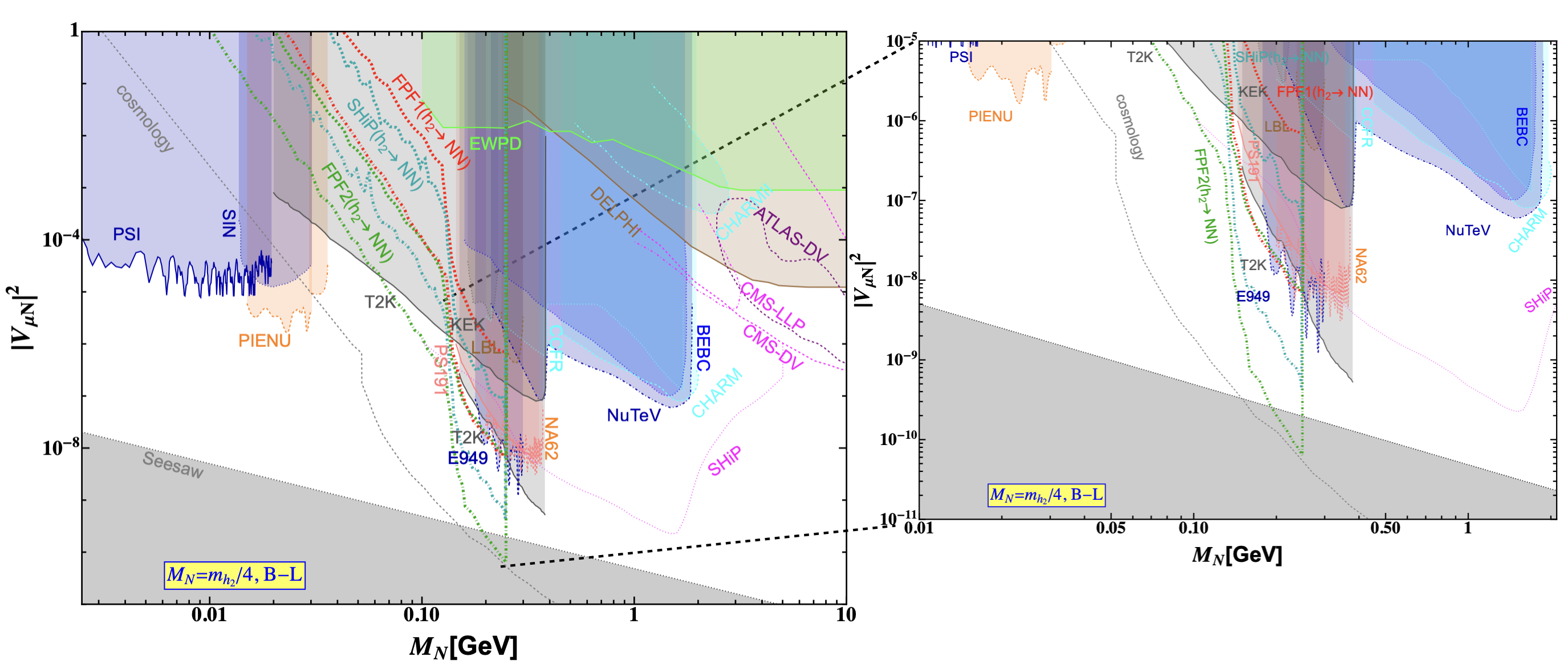}
\caption{Prospective limits on $|V_{e(\mu) N}|^2$ in the upper (lower) panel as a function of $M_N$ for long-lived heavy neutrinos produced via the decay of a short-lived scalar, assuming $\sin\theta = 10^{-3}$ (dot-dashed) and $\sin\theta=10^{-4}$ (dashed). The red, dark-green and dark-cyan curves correspond to FPF1, FPF2 and SHiP experiments considering $M_{N} = m_{h_2}/4$ for the B$-$L scenario. The shaded regions are excluded by existing experimental constraints and theoretical estimations. In the right of each figure we magnify the relevant region which could be probed by FPF and SHiP experiments.}
\label{mix2}
\end{figure*}
%%%%%%%%%%%%%%%%%%%%%%%%%%%%%%%%%%%%%%%%%%%%%%%%%%%%%%%%%%%%%%%%%%%%%%%%%%%%%%%%%%%
We explore the sensitivity of FPF@FCC and SHiP in our work. For that we need to explore the production of the BSM scalar $h_2$ in these experiments. The major source of the BSM scalars should be the the decay of mesons through the mixing of BSM scalars to the SM higgs. As the coupling of a meson to the SM higgs is proportional to its mass, we have to focus on heavier mesons with significant abundance from the initial proton-proton collision or proton collision with the target nucleus. Therefore, the best choice for $h_2$ production is through the two and three body decays of B mesons following
$B^\pm \rightarrow K^\pm + h_2$, $B^0 (\bar{B^0}) \rightarrow K_L^0 (\bar{K_L^0}) + h_2$, $B^\pm \rightarrow K^\pm + \gamma + h_2$ and $B^0 (\bar{B^0}) \rightarrow K_L^0 (\bar{K_L^0}) + \gamma + h_2$, respectively. In the following analysis, we ignore contributions from kaon decays since they are relevant only in parameter regions already excluded by current experimental constraints. In addition, scalar production from $D$-meson decays is strongly suppressed because the corresponding process is not mediated by top-quark loops.
The differential cross-section of the the production of $h_2$ from 2-body and 3-body decays would be,
\small
\bea
\label{eq:width-NN}
\frac{d\sigma_{h_2}}{dp_{h_2}d\cos\theta_{h_2}} &=& \frac{d\sigma_M}{dp_Md\cos\theta_M} \text{BR}(M\rightarrow h_2 + X)   \\ %\text{for 2-body decay}\\
\frac{d\sigma_{h_2}}{dp_{h_2}d\cos\theta_{h_2}} &=& \frac{d\sigma_M}{dp_Md\cos\theta_M} \times \nonumber \\
&&\text{BR}(M\rightarrow h_2 +  \gamma + X)~~~~~  %\text{for 3-body decay}
\eea
\normalsize
and the corresponding branching ratios are \cite{Chivukula:1988gp,Grinstein:1988yu,Feng:2017vli,FASER:2018eoc,FASER:2019aik}  %for of production $h_2$ from meson $M$ is,
\begin{align}
&\text{BR}(M\rightarrow h_2 + X) = 5.6  \sin^2\theta  \left(1- \frac{m_{h_2}^2}{m_b^2}\right)^2   \\% \text{for 2-body decay}
&\frac{d\text{ BR}(M\rightarrow h_2 +\gamma+ X)}{{dq^2d\cos\theta}} = 7.37\times 10^{-10} \left(\frac{2m_{h_2}}{9}\right)^3 \times \nonumber \\
&\hspace{3.5cm}\sqrt{1-4\frac{m_{h_2}^2}{q^2}} \left(1-\frac{q^2}{m_b^2} \right)^2~~~ %&& \text{for 3-body decay}
\end{align}
where $m_b$ is the $b$-quark mass and $q^2$ is the total momentum of $h_2$ and $\gamma$ in the center of mass frame.

In our work, the scalar $h_2$ is considered to be a short-lived particle which quickly decays dominantly into a pair of long-lived RHNs. The long-lived RHNs fly to the detector and decay into visible signal comprising of SM charged leptons and hadrons inside the decay volume of FPF and SHiP experiments. The number of detected signal events would be
%%%%%%%%%%%%%%%%%%%%%%%%%%%%%%%%%%%%%%%%%%%%%%%%%%
\begin{align}
&N_\text{signal} (m_{h_2}, \sin\theta) = \frac{d\sigma_{h_2}}{dp_{h_2}d\cos\theta_{h_2}}  2  \text{ BR} (h_2\rightarrow NN) \nonumber \\
&\text{ Acc}(\vec{p}_N, M_N,|V_{N\ell}|^2)
\text{ BR}(N \rightarrow \text{visible})
\end{align}
%%%%%%%%%%%%%%%%%%%%%%%%
where acceptance of the RHNs by the detector can be defined as
\begin{align}
&\text{Acc}(\vec{p}_N, M_N,|V_{N\ell}|^2) = \mathcal{P}(\vec{p}_N, M_N,|V_{N\ell}|^2)\times\nonumber\\
&\hspace{4cm}\text{BR} (N\to \text{visible})~.
\label{eq:accep_0}
\end{align}
where
\begin{align}
\mathcal{P}(\vec{p}_N, M_N,|V_{N\ell}|^2)&=(e^{-L_1/d_{\rm lab}} - e^{-(L_1 + L_2)/d_{\rm lab}}) \times \nonumber\\
&\hspace{1cm}F(X, Y)\nonumber \\
&\approx \frac{L_2}{d_{\rm lab}}e^{-L_1/d_{\rm lab}} F(X, Y).
\end{align}
%%%%%%%%%%%%%%%%%%%%%%%%%%%%%%%%%%%%%%%%%%%%%%
The first term in the brackets represents the probability that the RHNs decays within the detector volume, namely between $L_1$ and $L_1+L_2$~\cite{A:2025ygb,KA:2026czk}. Here, $d_{\rm lab}$ is the RHNs decay length in the laboratory frame, given by $d_{\rm lab}=c\tau\,\frac{p_N}{M_{N}}$,
where $\tau$ is the proper lifetime of the RHNs. The angular acceptance of the emitted RHNs is determined by the factor $F(X,Y)$ where $X$ and $Y$ are the lengths of the cross-sectional area of the detector~\cite{KA:2026czk}. The specifications of the FPF and SHiP detectors are given in Tab.~\ref{tab:placeholder}. In our analysis, we consider the mass range $0.01~\text{GeV}<m_{h_2}<1~\text{GeV}$ and fix $M_N=m_{h_2}/4$ to derive sensitivities to the scalar mixing angle as a function of $m_{h_2}$ and to the light-heavy neutrino mixing as a function of the heavy neutrino mass. We restrict our study to $m_{h_2}<1$ GeV to avoid the large theoretical uncertainties associated with modeling hadronic decay widths in the few-GeV mass region. The projected sensitivities are obtained assuming a conservative background-free analysis with 3 signal events corresponding to a $95\%$ CL sensitivity.

Considering two different benchmarks of neutrino mixing $|V_{e(\mu) N}|^2=10^{-6}$ (solid) and $10^{-8}$ (dashed) we estimate prospective bounds on the scalar mixing angle $\sin\theta$ as a function of $m_{h_2}$ where short-lived scalar decays into a pair of long-lived heavy neutrinos which decay into visible modes comprising of charged leptons and hadrons. These bounds are shown in the left (right) panel of Fig.~\ref{mix1} for FPF1 (red), FPF2 (green) and SHiP (blue) experiments considering heavy neutrinos dominantly decays into electrons (muons). We obtain the strongest constraints on the scalar mixing angle in the mass range $0.5~\mathrm{GeV} \leq m_{h_2} \leq 1~\mathrm{GeV}$. In this region, we find that SHiP can probe significantly smaller scalar mixing angles than FPF2, improving the sensitivity by approximately two orders of magnitude. This enhancement is primarily due to the large projected PoT, $6\times10^{20}$, expected at SHiP. We compare our results with the existing excluded region, shown in dark gray, as well as the projected sensitivities of HIKE~\cite{HIKE:2022qra}, DUNE~\cite{Berryman:2019dme}, CODEX-b~\cite{CODEX-b:2019jve}, MATHUSLA~\cite{Bertuzzo:2020rzo,Curtin:2018mvb}, and SHiP~\cite{SHiP:2015vad}. It should be noted, however, that these existing and projected limits are derived under the assumption that the singlet scalar decays exclusively into SM particles, i.e., in scenarios where the SM is extended only by a singlet scalar without heavy neutrinos. Consequently, they are not directly applicable to the scenario considered in this work, where the scalar predominantly decays into heavy neutrinos. Finally, we emphasize that the bounds shown in Fig.~\ref{mix1} are highly sensitive to the light-heavy neutrino mixing. For sufficiently small mixing, the RHNs become too long-lived and predominantly decay outside the detector, whereas for sufficiently large mixing, they decay before reaching the detector. In both cases, the number of detectable signal events is significantly reduced, leading to weaker constraints.

Next, we consider two benchmark values of the scalar mixing angle, $\sin\theta=10^{-3}$ (dot-dashed) and $\sin\theta=10^{-4}$ (dashed), and estimate the prospective limits on the light-heavy neutrino mixing, $|V_{\ell N}|^2$, as a function of the heavy neutrino mass, $M_N$. We consider two scenarios in which the heavy neutrino predominantly decays into the electron or muon channel, with the corresponding results shown in the upper and lower panels of Fig.~\ref{mix2}, respectively. The projected sensitivities of FPF1, FPF2, and SHiP are indicated by the red, dark-green, and dark-cyan curves. The right-hand side of each panel provides a magnified view of the parameter region accessible to these future experiments. For $\sin\theta=10^{-4}$, the projected sensitivities of FPF1, FPF2, and SHiP to $|V_{eN}|^2$ remain weaker than the current experimental limits. In contrast, for $\sin\theta=10^{-3}$, FPF2 and SHiP can improve the existing constraints, reaching sensitivities in the range $3\times10^{-10}\lesssim |V_{eN}|^2 \lesssim 2\times10^{-7}$ for $0.125~\mathrm{GeV}\leq M_N\leq0.25~\mathrm{GeV}$, thereby surpassing all current bounds in this mass region.
For the muon channel, when $\sin\theta=10^{-3}$, the projected sensitivities of FPF1, FPF2, and SHiP span $3\times10^{-10}\lesssim |V_{\mu N}|^2 \lesssim 10^{-5}$ over the mass range $0.07~\mathrm{GeV}\leq M_N\leq0.25~\mathrm{GeV}$, exceeding the existing experimental constraints. For the smaller scalar mixing, $\sin\theta=10^{-4}$, FPF2 and SHiP remain sensitive to $10^{-8}\lesssim |V_{\mu N}|^2 \lesssim 10^{-6}$ in the range $0.15~\mathrm{GeV}\leq M_N\leq0.25~\mathrm{GeV}$, providing stronger limits than those currently available.

We compare our projected sensitivities with the existing experimental constraints on the light-heavy neutrino mixing, $|V_{eN}|^2$. At low masses, the strongest limits arise from peak searches in pion decays at TRIUMF \cite{Britton:1992xv} and PIENU \cite{PIENU:2017wbj,Bryman:2019bjg,PIENU:2019usb} for $M_N\lesssim0.15~\mathrm{GeV}$, while Borexino \cite{Borexino:2013bot} provides the leading constraints in the range $0.003~\mathrm{GeV}\leq M_N\leq0.016~\mathrm{GeV}$ through the decay $N\to\nu e^+e^-$. Above the pion threshold, NA62 \cite{NA62:2020mcv,NA62:2025csa} sets the strongest bounds from the peak search in $K^+\to e^+N$ for $0.15~\mathrm{GeV}\leq M_N\leq0.45~\mathrm{GeV}$, while the T2K near detector \cite{T2K:2019jwa} probes the region $0.4~\mathrm{GeV}\leq M_N\leq0.5~\mathrm{GeV}$. At higher masses, beam-dump experiments such as BEBC \cite{WA66:1985mfx,Barouki:2022bkt} and CHARM \cite{CHARM:1985nku} constrain heavy neutrinos up to $M_N\simeq2.25~\mathrm{GeV}$, whereas Belle \cite{Zhou:2021ylt}, DELPHI \cite{DELPHI:1996qcc}, ATLAS \cite{Tastet:2021vwp,ATLAS:2022atq}, and CMS \cite{CMS:2022fut} extend the exclusion reach up to $M_N\simeq16~\mathrm{GeV}$. We also compare our results with displaced-vertex searches from NA62 \cite{NA62:2020mcv}, ATLAS \cite{ATLAS:2019kpx,ATLAS:2022atq,ATLAS:2024fdw}, CMS \cite{CMS:2024xdq,CMS:2024bni}, and the projected sensitivity of the SHiP experiment \cite{Alekhin:2015byh,SHiP:2018xqw}. These constraints are shown in the upper panel of Fig.~\ref{mix2}.

Similarly, we compare our projected sensitivities with the existing constraints on $|V_{\mu N}|^2$. At low masses, PSI \cite{Daum:1987bg} and PIENU \cite{PIENU:2019usb} constrain heavy neutrinos produced in pion decays for $1.2\times10^{-3}~\mathrm{GeV}\leq M_N\leq0.025~\mathrm{GeV}$, while the T2K near detector \cite{Arguelles:2021dqn} provides complementary limits from heavy neutrino decays in flight. At higher masses, stringent bounds are obtained from KEK \cite{Hayano:1982wu}, MicroBooNe \cite{Kelly:2021xbv}, BNL \cite{BNL-E949:2009dza}, BEBC \cite{WA66:1985mfx}, NuTeV \cite{NuTeV:1999kej}, CHARM \cite{CHARM:1985nku}, NA62 \cite{NA62:2025csa}, ATLAS \cite{ATLAS:2019kpx,ATLAS:2022atq}, and CMS \cite{CMS:2022fut}. We also compare our results with the projected sensitivity of the SHiP experiment \cite{SHiP:2018xqw}. These constraints are shown in the lower panel of Fig.~\ref{mix2}.

In both panels of Fig.~\ref{mix2}, we also include the indirect cosmological constraints from CMB and BBN \cite{Vincent:2014rja,Dolgov:2000pj,Ruchayskiy:2012si,Gelmini:2020ekg,Langhoff:2022bij,Sabti:2020yrt,Boyarsky:2020dzc}, shown by the gray dot-dashed curve, which constrain $|V_{e(\mu)N}|^2$ for $M_N\lesssim0.2~\mathrm{GeV}$. For comparison, we also show the theoretical lower bound from the seesaw relation, assuming $m_\nu\leq0.1~\mathrm{eV}$, indicated by the gray dashed curve labeled ``Seesaw''.

%%%%%%%%%%%%%%%%%%%%%%%%%%%%%%%%%%%%%%%%%%%%%%%%%%%%%%%
\section{Conclusions}
\label{secIV}
%%%%%%%%%%%%%%%%%%%%%
In this work, we have investigated the prospects of probing a light singlet scalar and long-lived heavy neutral leptons at the proposed FPF@FCC and SHiP experiments within the minimal $U(1)_{B-L}$ framework. Unlike conventional LLP scenarios, where the scalar itself is long-lived and decays into SM particles, we considered a complementary scenario in which a light scalar produced in rare meson decays promptly decays into a pair of long-lived heavy neutrinos. These heavy neutrinos subsequently travel macroscopic distances before decaying into visible charged leptons and hadrons inside the detector.

Taking into account realistic detector geometries, decay probabilities, meson production rates, and visible branching fractions, we estimated the projected sensitivities of FPF1, FPF2, and SHiP through a conservative background-free analysis corresponding to three signal events at the $95\%$ confidence level. We studied the sensitivity to the scalar-Higgs mixing angle as a function of the scalar mass and to the light-heavy neutrino mixing as a function of the heavy neutrino mass for scenarios in which the heavy neutrino predominantly mixes with either the electron or muon flavor.

For the scalar mixing, we find that SHiP provides the strongest sensitivity in the mass range $0.5~\mathrm{GeV}\lesssim m_{h_2}\lesssim1~\mathrm{GeV}$, improving upon the reach of FPF2 by approximately two orders of magnitude owing to its large projected proton statistics. We also emphasize that these sensitivities differ significantly from those obtained in conventional singlet-scalar searches because, in our scenario, the scalar predominantly decays into heavy neutrinos rather than directly into SM particles. Furthermore, the projected limits depend strongly on the light-heavy neutrino mixing, since excessively small (large) mixing causes the heavy neutrinos to decay outside the detector (before reaching the detector), thereby reducing the observable signal.

For the light-heavy neutrino mixing, we find that future intensity-frontier experiments can substantially improve the existing bounds in the low-mass region. In particular, for $\sin\theta=10^{-3}$, FPF2 and SHiP can surpass current constraints on $|V_{eN}|^2$ in the mass range $0.125~\mathrm{GeV}\leq M_N\leq0.25~\mathrm{GeV}$, while for the muon channel both FPF2 and SHiP improve the present limits over a broad region of parameter space, with FPF1 also providing competitive sensitivity for larger scalar mixing.

Our results demonstrate that the proposed FPF@FCC and SHiP experiments offer powerful and complementary probes of light scalars and long-lived heavy neutral leptons. They significantly extend the discovery potential for the scalar and neutrino sectors of the minimal $U(1)_{B-L}$ model beyond existing searches and provide a promising avenue for testing the origin of neutrino masses through long-lived particle signatures at future intensity-frontier facilities.

%%%%%%%%%%%%%%%%%%%%%%%%%%%%%%%%%
\begin{widetext}
\section*{Acknowledgements}
S.K.A is supported by JST SPRING, Grant Number JPMJSP2119. The work of S.M. is supported by KIAS
Individual Grants (PG086002) at Korea Institute for Advanced Study.
\appendix
\label{appendix}
%%%%%%%%%%%%%%%
\section{Relevant partial decay modes of the BSM scalar }
\label{sec:decays1}
%%%%%%%%%%%%%%%%%%%%%
In the following we summarize the BSM scalar decay channels relevant to our analysis:
\begin{align}
\label{eqn:phiZpZp}
& \Gamma(h_2 \to f{\bar f}) = \sin^2{\alpha} \times \frac{ N_c m_{h_2} m_f^2}{8\pi v_{h}^2} \left( 1- \frac{4m_f^2}{m_{h_2}^2} \right)^{\frac32}, \\
& \Gamma(h_2 \to \pi^+\pi^-) = 2\Gamma(h_2 \to\pi^0\pi^0) = \frac{G_F m_{h_2}^3 \sin^2\alpha}{8\sqrt{2}\pi} \left( 1 - \frac{4m_\pi^2}{m_{h_2}^2} \right)^{1/2} \left| G(m_{h_2}^2) \right|^2 ,\\
& \Gamma (h_2 \to Z'Z')  =  c_\alpha^2\frac{g_{BL}^2 m_{h_2}^3}{8 \pi M_{Z'}^2}
\left( 1- \frac{4 M_{Z'}^2}{m_{h_2}^2} \right)^{1/2}
\left( 1- \frac{4 M_{Z'}^2}{m_{h_2}^2} + \frac{12 M_{Z'}^4}{m_{h_2}^4} \right) \,,\\
& \Gamma(h_2 \to N_\alpha N_\alpha) = \frac{Y_{N_\alpha}^2 m_{h_2} \cos^2\alpha}{32 \pi}  \Big(1-4\frac{M_{N_\alpha}^2}{m_{h_2}^2}\Big)^{3/2}, \\
& \Gamma (h_2 \to \gamma\gamma) = \frac{G_F \alpha_e^2 m_{h_2}^3 \sin^2\alpha}{128 \sqrt{2}\pi^3} \left| \sum_f N_c Q_f^2 A_{1/2} (\tau_f) + A_1(\tau_W) \right|^2
\label{h2photon}
\end{align}
where  $f$ denotes a fermion of single generation and $Q_f$ denotes the corresponding fermionic charge in unit of a proton charge. In our analysis we consider a parameter region where $M_{Z^\prime} > m_{h_2}$ disallowing $Z^\prime$ pair production from the decay of $h_2$. Among other expressions $\alpha_e (= 1/137)$ is the electromagnetic fine structure constant, $G_F(=1.1663787\times 10^{-5} {\rm  GeV^{-2}})$ is the Fermi constant, and $v_h(= 246 {\rm GeV})$ is the electroweak VEV. Here $G(s)$ with $s=m_{h_2}^2$ is the amplitude of quark-gluon transition amplitude to pions, can be expressed in terms of three form factors as mentioned in \cite{Donoghue:1990xh,Berryman:2019dme},
\begin{equation}
\label{eqn:G}
    G(s) = \frac{2}{9} \theta_\pi(s) + \frac{7}{9} \Gamma_\pi(s) + \frac{7}{9}\Delta_\pi(s).
\end{equation}
The corresponding form factors in chiral perturbation theory are expressed below as
\begin{align}
\label{eqn:piff}
&\theta_\pi(s) = \left( 1 + \frac{2m_\pi^2}{s} \right) (1+\psi(s)) + b_\theta s, \\
&\Gamma_\pi(s) = \frac{m_\pi^2}{s}(1+\psi(s)+b_\Gamma s) , \\
&\Delta_\pi(s) = d_F(1+\psi(s)+b_\Delta s)\\
\end{align}
where the constants are given as $b_\theta = 2.7 \text{GeV} ^{-2}$, $b_\Gamma = 2.6 \text{GeV} ^{-2}$, $b_\Delta = 3.3 \text{GeV} ^{-2}$, $d_F = 0.09$, and the function is given by
\begin{equation}
    \label{eqn:psi}
    \psi(s) = \frac{2s-m_\pi^2}{16\pi^2 f_{\pi}^2} \left[ \kappa \log\left(\frac{1-\kappa}{1+\kappa}\right) + 2 + i\pi\kappa\right] + \frac{s}{96\pi^2 f_{\pi}^2}
\end{equation}
with $f_\pi = 130 \text{MeV}$ as the pion decay constant, and $\kappa=\sqrt{1-4m_\pi^2/s}$. In Eq.~(\ref{h2photon}) we used $A_{1/2}(\tau_f)$ to denote a fermionic loop function, and $A_1(\tau_W)$ to denote a $W$ boson loop, respectively. They are defined below \cite{Djouadi:2005gi,Berryman:2019dme} as
\begin{align}
\label{eqn:Wfloops}
&A_{1/2}(\tau) = \frac{2}{\tau^2}[\tau + (\tau-1)f(\tau)], \\
&A_1(\tau) = - \frac{1}{\tau^2}[2\tau^2 + 3\tau + 3(2\tau-1)f(\tau)]
\end{align}
with $\tau_X = m_{h_2}^2/4m_X^2$, and
\begin{equation}
\label{eqn:fstep}
    f(\tau)= \begin{cases}
\arcsin^2\sqrt{\tau}, & \tau \le 1 \\
-\frac{1}{4} \left[ \log \left(\frac{1 + \sqrt{1 - 1/\tau}}{1 -\sqrt{1 - 1/\tau}} \right) - i\pi \right]^2, & \tau > 1
\end{cases}
\end{equation}
\end{widetext}
%%%%%%%%%%%%%%%%%%%%%%%%%%%%%%%%%%%%%%%%
%\vspace{-0.398in}
\bibliographystyle{utphys}
\bibliography{bibliography}
%%%%%%%%%%%%%%%%%%%%%%%%%%%%%%%%%%%%%%%%%%%%%
\end{document}